\documentclass[aps,prl,twocolumn]{revtex4-2}
\usepackage{graphicx} % Required for inserting images
\usepackage{amsmath}
\usepackage{amsfonts}
\usepackage{amssymb}
\usepackage{amsbsy}
\usepackage{color}
\usepackage{ulem}
\usepackage{tikz}
\usepackage{dsfont}
\usepackage{multirow}

\newcommand{\beq}{\begin{equation}}
\newcommand{\eeq}{\end{equation}}
\newcommand{\bea}{\begin{eqnarray}}
\newcommand{\eea}{\end{eqnarray}}

\newcommand{\ham}{\mathcal{H}}
\newcommand{\uevol}{\mathcal{U}}

\begin{document}

\title{Destructive Interference induced constraints in Floquet systems}
\author{Somsubhra Ghosh$^1$, Indranil Paul$^2$,
K. Sengupta$^3$, Lev Vidmar${^{1,4}}$}
\affiliation{$^1$Department of Theoretical Physics, J. Stefan Institute, SI-1000 Ljubljana, Slovenia. \\
$^2$Universit\'{e} Paris Cit\'{e}, CNRS, Laboratoire Mat\'{e}riaux
et Ph\'{e}nom\`{e}nes Quantiques, 75205 Paris, France.\\
$^3$School of Physical Sciences, Indian Association for the Cultivation of Science, Kolkata 700032, India.\\
$^4$Department of Physics, Faculty of Mathematics and Physics, University of Ljubljana, SI-1000 Ljubljana, Slovenia\looseness=-1}
\date{\today}

\begin{abstract}
We introduce the paradigm of destructive many-body interference between
quantum trajectories as a means to systematically generate prethermal kinetically constrained dynamics
in Floquet systems driven at special frequencies. Depending on the processes that are suppressed by interference, the constraint may or may not be associated with an emergent global conservation; the latter kind having no mechanism of generation in time-independent settings. As an example, we construct an one-dimensional interacting spin model exhibiting strong Hilbert space fragmentation with and without dipole moment conservation, depending on the drive frequency. By probing the spatiotemporal profile of the out-of-time-ordered correlator, we show that this model, in particular, has initial states in which quantum information can be spatially localized - a useful feature in the field of quantum technologies. Our paradigm unifies various types of Hilbert space fragmentation that can be realized in driven systems. 

\end{abstract}

%\pacs{03.75.Lm, 05.30.Jp, 05.30.Rt}

\maketitle 

{\it Introduction --}
Kinetically constrained systems have long been a breeding ground of exotic non-ergodic phenomena ranging from Hilbert space fragmentation (HSF) and localization to quantum many-body scars and dynamical freezing \cite{hsf1,hsf2,hsf3,hsf4,hsf5,hsf6,hsf7,hsf8,hsf9,hsf10,kc1,kc2,kc3,kc4,scarref1,scarref2,scarref3,scarref4,flscar1,flscar2,flscar3,flscar4,flscar5,adas1,adas2,dsen1,luitz1,dsen2}. Strong HSF, in particular, leads to strong ergodicity breaking by confining a typical initial state within an exponentially small fragment of the Hilbert space \cite{hsf1,hsf2}. However, in most cases, the constraints are imposed {\it a priori} on the systems, and little discussion \cite{hsf5,dsen2,sghosh1} exists on how to generate them organically, starting from an unconstrained Hamiltonian system. 

The constraints which have been studied in the literature so far can broadly be classified into two categories, which we dub here as Type-I and Type-II constraints. Type-I constraints are those which are associated with a global conservation law, for instance, dipole moment conservation \cite{hsf1,hsf2} or conservation of number of nearest-neighbor particle pairs \cite{hsf4,hsf5,sghosh1}. Type-II constraints, on the other hand, cannot be readily associated with any globally conserved quantity. Examples of this type include the constraint in the number-conserving East model or the third-order Floquet Hamiltonian in the driven PXP model \cite{hsf8}. Note that by global quantities, we refer to quantities which can be expressed as sums of local operators with the sum being taken over the full system. 

Type-I constraints can be generated (at least approximately) in time-independent setting by imposing a high energy cost on moves which violate the constraints \cite{hsf4,hsf5}. Recently, some studies have also extended this idea to realize prethermal HSF in non-equilibrium Floquet systems \cite{sghosh1,dsen2}. For Type-II constraints, a Floquet version of the number non-conserving quantum East Model has been studied using unitary gates \cite{uc1} (this version does not exhibit strong HSF). However, to the best of our knowledge, no such mechanism exists to systematically generate Type-II constraints in time-independent and time-dependent Hamiltonian settings starting from an unconstrained system. 
\begin{figure}[!t]
    \centering
    \includegraphics[width=0.95\linewidth]{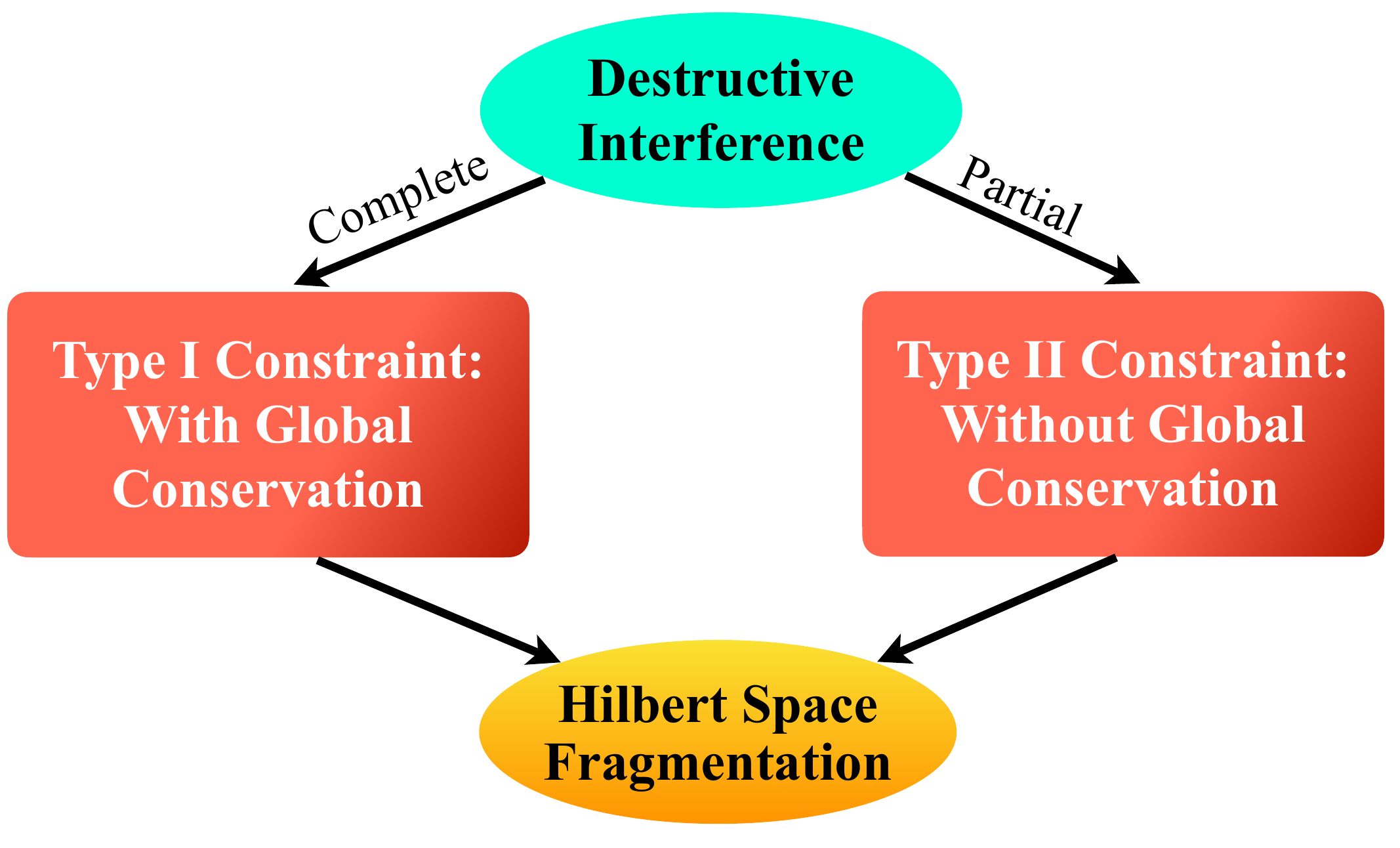}
    \caption{Schematic showing two types of constraints which appear in the study of HSF. Our study proposes that destructive interference provides a route to realize both types of constraints, depending on whether the interference is `complete' or `partial'. See text for details.}
    \label{fig:fig1}
\end{figure}

In this Letter, we propose a general framework for generating both kinds of prethermal kinetic constraints in Hamiltonian Floquet systems by utilizing the principle of destructive many-body interference between quantum trajectories within one drive period. Depending on whether the destructive interference is `complete' or `partial' (as explained below), it is possible to enforce Type-I or Type-II constraints, respectively, at \textit{stroboscopic time steps} by driving the system at special frequencies. Our study thus provides the first step towards a unifying approach to generating kinetic constraints and realizing HSF in driven systems (Fig. \ref{fig:fig1}). 

As a particular example, we consider a periodically driven one-dimensional (1-d) spin-$1/2$ chain which exhibits prethermal strong HSF with or without dipole moment conservation, depending on the chosen drive frequency. In each of these cases, the study of the out-of-time-ordered correlator (OTOC) shows that for special initial states, information can be localized spatially within the spin chain. This opens up a new avenue to realize HSF-induced spatial localization of quantum information in periodically driven interacting systems.

{\it General Framework --} We consider short-range, time-periodic Hamiltonians of the form $\ham(t)=\ham_0(t)+\ham_1$, where
\bea
    \ham_0(t)&=& V(t)\hat{Q}\equiv V(t)\sum_j \hat{Q}_j; \quad \ham_1=\sum_j J_j \hat{P}_j.
    \label{eq:ham_gen}
\eea
Without loss of generality, we assume that $\hat{Q}_j$ is diagonal in the computational basis and $\hat{P}_j$ generates quantum fluctuations (hoppings, spin-flips {\it et cetera}) in the system, with $[\hat{P}_i,\hat{Q}_j]\neq 0$ in general. Here, $j$ is a collective index that refers to the sites over which the corresponding operator, $\hat{P}_j$ or $\hat{Q}_j$ has support. The energy parameter $V(t)$ is driven periodically with a time period $T=2\pi/\omega_D$ and a drive amplitude $V_0\gg J_j$, for all $j$'s, so that the Floquet Hamiltonian $\ham_F$ can be perturbatively estimated using the Floquet Perturbation Theory (FPT) or, equivalently, the Rotating Wave Approximation (RWA) \cite{fl1,fl2,fl3}.

{\it Quantum Trajectories --} The exact Floquet unitary operator $\uevol_F$ encodes the time evolution of the system within one complete drive period $0 \to T$. In particular, for a transition like $|\psi(0)\rangle=|m\rangle\to |\psi(T)\rangle=|n\rangle$, the net transition amplitude $\langle n|\uevol_F|m\rangle$ is obtained by summing over all possible trajectories in the time domain which take the system from $|m\rangle$ at time $t=0$ to state $|n\rangle$ at $T$. This follows from the quantum superposition principle. Note, for future reference, that both $|m\rangle$ and $|n\rangle$ are computational basis states. 

\begin{figure}
    \centering
    \includegraphics[width=0.95\linewidth]{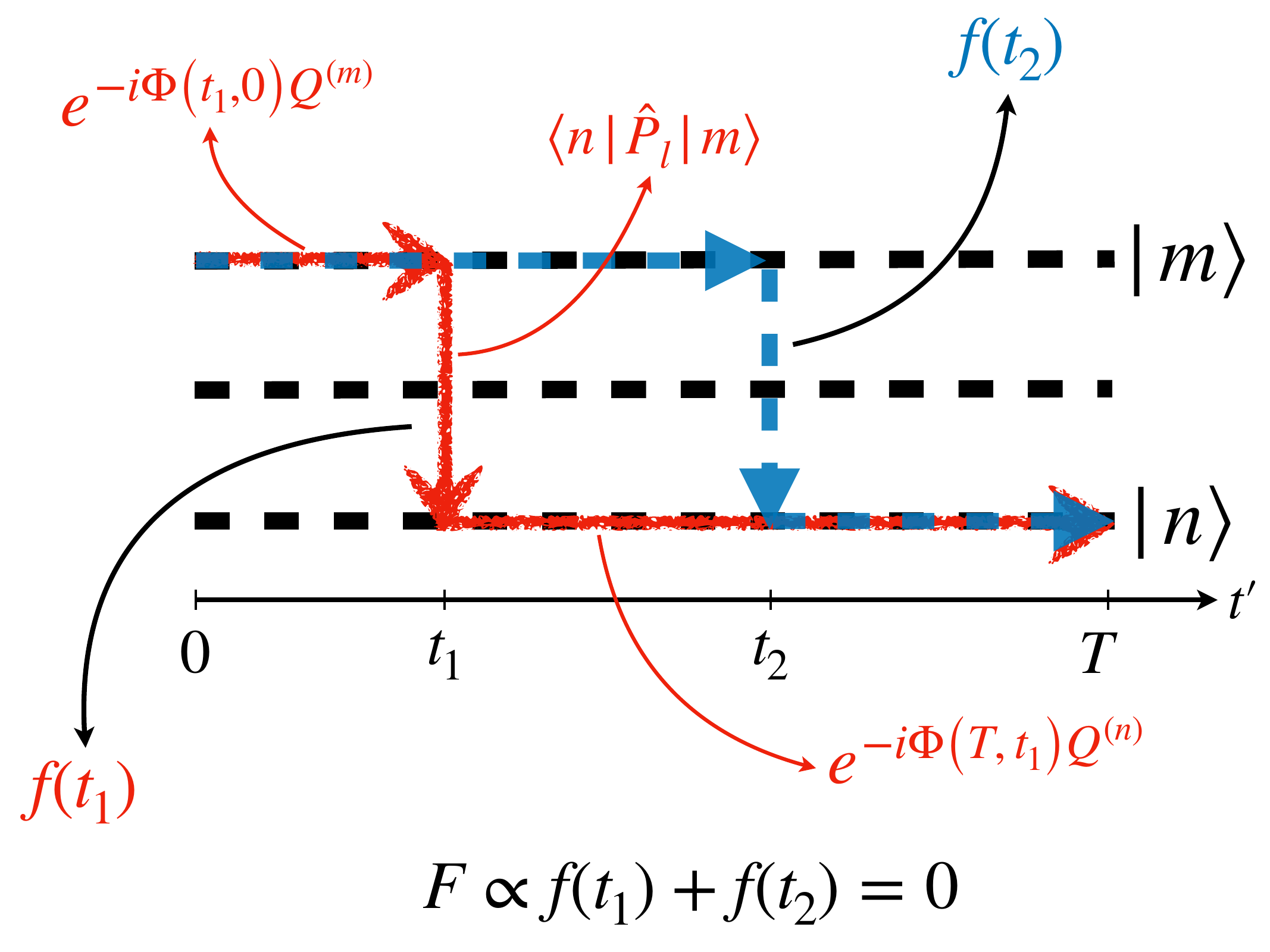}
    \caption{(Color online) Two representative quantum trajectories for the first-order transition $|\psi(0)\rangle=|m\rangle\to|\psi(T)\rangle=|n\rangle$, where $|n\rangle=\hat{P}_l|m\rangle$. The trajectory in red is described as follows: From time $0\to t_1$, the system stays at $|m\rangle$, accumulating a phase $e^{-i\Phi(t_1,0)Q^{(m)}}$ due to $\ham_0$; at time $t_1$, the system jumps from $|m\rangle\to|n\rangle$ under the action of $\ham_1$; and finally for the rest $t_1\to T$, it stays at $|n\rangle$ gathering a phase $e^{-i\Phi(T,t_1)Q^{(n)}}$. The total phase factor associated with this trajectory is thus given by $f(t_1)$ in Eq. (\ref{eq:phaselag}). A similar description exists for the blue trajectory with $t_1\leftrightarrow t_2$. If, at the chosen frequency, these two trajectories interfere destructively, then their net contribution to the transition amplitude is $F=0$. If $\omega_D$ is such that all trajectories corresponding to $\langle n|\uevol_F^{(1)}|m\rangle$ interfere destructively, then this transition is suppressed in first-order. See text for details.}
    \label{fig:fig2}
\end{figure}

When $\uevol_F$ is expanded in a perturbative series using FPT, the $k$-th order term, $\uevol_F^{(k)}$, having $k$ powers of $\ham_1$, facilitates this transition through $k$ jumps via $(k-1)$ intermediate states (see End Matter and \cite{SI}). In between any two successive jumps, the system evolves under $\ham_0$ gathering a phase factor, as we explain next. 

A typical quantum trajectory for the first-order approximation of this transition is depicted in red in Fig. \ref{fig:fig2}. From time $t'=0\to t_1$, the state $|m\rangle$ evolves only under $\ham_0$ gathering a phase $e^{-i\Phi(t_1,0)Q^{(m)}}$, where $\hat{Q}|m\rangle=Q^{(m)}|m\rangle$ and $\Phi(t_1,0)=\frac{1}{\hbar}\int_0^{t_1} dt' V(t')$. At time $t'=t_1$, there is a jump from $|m\rangle \to |n\rangle$ under the action of $\ham_1$ and finally from $t'=t_1\to T$, the state stays at $|n\rangle$ accumulating a phase $e^{-i\Phi(T,t_1)Q^{(n)}}$. This trajectory therefore carries a net phase of
\bea
    f(t_1)&=&\exp{\left(-i\left[\Phi(T,t_1)Q^{(n)}+\Phi(t_1,0)Q^{(m)} \right] \right)}\nonumber\\
    &=& \exp{\left(i\Phi(t_1,0)\Delta Q^{(nm)} \right)}.
    \label{eq:phaselag}
\eea
The second line follows under the assumption $\Phi(T,0)=0$, which is most often the case for commonly used drive protocols, and $\Delta Q^{(nm)}=Q^{(n)}-Q^{(m)}$. If $|n\rangle$ is connected to $|m\rangle$ by the process $\hat{P}_l$, i.e., $|n\rangle\sim\hat{P}_l|m\rangle$, then the contribution of this trajectory to the transition amplitude is $f(t_1)J_l\langle n|\hat{P}_l|m\rangle$. It is useful to note here that $Q^{(m)}$ and $Q^{(n)}$, being sums of few-body operators, differ from each other only in the region of space which overlaps with $\hat{P}_l$. Thus, $\Delta Q^{(nm)}\equiv \Delta Q_l^{(nm)}$ \cite{DQevaluation}.

The intermediate time step $t$ when the jump from $|m\rangle\to|n\rangle$ occurs can lie anywhere between $0\to T$, and one needs to sum over all possible values of the phases $f(t)$ to obtain the total amplitude (in first-order), giving a net contribution
\beq
    \langle n|\uevol_F^{(1)}|m\rangle =-\frac{iT}{\hbar} J_l F\left(\frac{V_0\Delta Q_l^{(nm)}}{\hbar\omega_D} \right)\langle n|\hat{P}_l|m\rangle,
    \label{eq:UF1mn}
\eeq
where $F\left(\frac{V_0\Delta Q^{(nm)}}{\hbar\omega_D} \right)=\frac{1}{T}\int_0^T dt f(t)$. 
Using the definition $\uevol_F^{(1)}=-\frac{i}{\hbar}\ham_F^{(1)}T$, this gives for the first-order Floquet Hamiltonian \cite{SI,DQevaluation}
\beq    
    \ham_F^{(1)}=\sum_j J_j F\left(\frac{V_0\Delta\hat{Q}_j}{\hbar\omega_D} \right) \hat{P}_j.
    \label{eq:HF1_gen}
\eeq

{\it Destructive Interference (DI) --} Equation (\ref{eq:HF1_gen}) shows that to first-order, the `hopping' amplitude $J_j$ is renormalized to $\Tilde{J}_j \equiv J_j F\left(\frac{V_0\Delta\hat{Q}_j}{\hbar\omega_D} \right)$ as a result of periodic driving. The renormalization factor $F$ depends on the drive frequency, $\omega_D$, the drive amplitude, $V_0$, and $\Delta \hat{Q}_j$. Imposing constraints on the dynamics would mean selectively allowing some processes $\hat{P}_j$'s while suppressing others by tuning the drive frequency to some special values $\omega_D=\omega^*$. 

For processes that conserve $\hat{Q}$ ($\Delta \hat{Q}_j=0$), every trajectory carries a phase $f(t)=1$, (see Eq. (\ref{eq:phaselag})), so that all of them interfere constructively over a period giving $F=1$. Such processes are always allowed. If the drive frequency is chosen such that $F$ vanishes identically for {\it all} possible $\Delta\hat{Q}_j\neq 0$, then we have Type-I constrained dynamics, in which the constraint is associated with the conservation of $\hat{Q}$. This means that all trajectories with $\Delta\hat{Q}_j\neq 0$ interfere destructively at this frequency, giving `complete' DI. On the other hand, if the chosen $\omega_D$ is such that $F$ vanishes for only a class of $\Delta \hat{Q}_j\neq 0$, while remaining non-zero for other $\Delta\hat{Q}_j\neq 0$, then we have Type-II constrained dynamics. This corresponds to `partial' DI when only a class of $\Delta\hat{Q}_j\neq 0$ trajectories interfere destructively. This imposes constraints on the processes allowed by $\ham_F^{(1)}$ without any obvious conservation law. Table \ref{table:classification} summarizes this result. The only requirement for this framework to be useful is to be able to write an energy operator $\hat{Q}$ (that can be expressed as a sum of local terms) whose expectation value changes by $\Delta Q$ under the action of the kinetic terms $\hat{P}_j$'s and a drive protocol, whose corresponding $F$ function has a common root for all those values of $\Delta Q$'s which need to be suppressed.

\begin{table}
\centering
\begin{ruledtabular}
\begin{tabular}{c|c}
    \textbf{Processes suppressed} & \textbf{Constraint type} \\ \hline
    {\it All} $\hat{P}_i$ with $\Delta \hat{Q}_i\neq 0$ & Type-I \\
    (Complete DI) & \\
    Some $\hat{P}_i$ with $\Delta \hat{Q}_i\neq 0$; others allowed & Type-II\\
    (Partial DI) & 
\end{tabular}
\end{ruledtabular}
\caption{Types of emergent constraints in $\ham_F^{(1)}$, see Eq.~(\ref{eq:HF1_gen}), depending on what kind of processes are suppressed at the chosen special frequency $\omega_D=\omega^*$. This involves vanishing of the amplitude $F$ for the concerned processes $\hat{P}_i$ at $\omega^*$.}
\label{table:classification}
\end{table}

{\it Spin-$1/2$ Model --} We consider a 1-d chain of spin-$1/2$'s involving spin-spin interaction between four nearest-neighbor spins and a periodically driven dipole moment term, with open boundary conditions. Following the notation of the previous sections, $\hat{Q}=\sum_{j=1}^L j\hat{S}_j^z \equiv \hat{p}$ and 
\bea
    \ham_1&=& \frac{J}{2}\sum_{j=2}^{L-2}[\hat{S}_{j-1}^+\hat{S}_j^-\hat{S}_{j+1}^-\hat{S}_{j+2}^+ + \hat{S}_{j-1}^+\hat{S}_j^-\hat{S}_{j+1}^+\hat{S}_{j+2}^-\nonumber\\
     &+& \hat{S}_{j-1}^+\hat{S}_j^+\hat{S}_{j+1}^-\hat{S}_{j+2}^- + \text{h.c.}], 
     \label{eq:spin_half_ham}
\eea
with $V(t)$ modulated using a square pulse protocol $V(t)=-V_0 \text{sgn}(\sin{\omega_D t})$. The instantaneous Hamiltonian $\ham(t)$ conserves the total $\hat{S}^z=\sum_j \hat{S}_j^z$ and we work in the regime where $V_0\gg J$. 
%so that the first-order Floquet Hamiltonian, $\ham_F^{(1)}$, can be perturbatively determined using FPT or RWA using $J/V_0$ as the perturbation parameter. 
This model is inspired from the dipole-conserving spin-$1$ model which was one of the first models to have studied the constrained dynamics leading to HSF, and HSF leading to spatial localization in time-independent setting \cite{hsf1,hsf2,hsf10}. Note that $\Delta p=0,\pm 2,\pm 4$ corresponding to the first, second and third terms of $\ham_1$ and their hermitian conjugates, respectively. 

\begin{table}[b]
\centering
\begin{ruledtabular}
\begin{tabular}{c|c|c}
    \textbf{Processes $\hat{P}_j$ in $\ham_1$} & $F$ & $\omega^*$ \\ \hline
     $\hat{S}_{j-1}^+\hat{S}_j^-\hat{S}_{j+1}^-\hat{S}_{j+2}^+$ & $1$ & Always allowed \\
    $\hat{S}_{j-1}^+\hat{S}_j^-\hat{S}_{j+1}^+\hat{S}_{j+2}^-$ &  $e^{i\gamma_0/2}\frac{\sin{(\gamma_0/2)}}{\gamma_0/2}$ & $\frac{V_0}{\hbar m}$ \\
    $\hat{S}_{j-1}^+\hat{S}_j^+\hat{S}_{j+1}^-\hat{S}_{j+2}^-$ & $e^{i\gamma_0}\frac{\sin{\gamma_0}}{\gamma_0}$ & $\frac{2V_0}{\hbar m}$
\end{tabular}
\end{ruledtabular}
\caption{Processes in $\ham_1$ along with the renormalization factor $F$, which results due to the periodic drive. The resulting $\ham_F^{(1)}$ is Eq.~(\ref{eq:HF1aspinhalf}). The right column lists the frequencies $\omega^*$ at which the renormalization factor $F$ (see Eq.~(\ref{eq:HF1_gen})) vanishes, and the trajectories corresponding to these processes undergo DI. Here, $\gamma_0=V_0T/\hbar$ and $m$ is an integer.}
\label{table:spin_model_F}
\end{table}

For the square pulse protocol, the phase factor in Eq.~(\ref{eq:phaselag}) associated with the trajectories are given by $f(t)=\Theta(T/2-t)e^{iV_0t\Delta p/\hbar}+\Theta(t-T/2)e^{iV_0(T-t)\Delta p/\hbar}$. Table \ref{table:spin_model_F} lists the renormalization factors $F$ associated with the processes in $\ham_1$, which are obtained by summing over these phase factors as in Eq. (\ref{eq:UF1mn}). Thus, the first-order Floquet Hamiltonian, obtained using FPT, reads
\bea
     \ham_F^{(1)}&=&\frac{J}{2}\sum_{j=2}^{L-2} \left[ \hat{S}_{j-1}^+\hat{S}_j^-\hat{S}_{j+1}^-\hat{S}_{j+2}^+ \right.\nonumber\\ 
     &+& \left. e^{i\gamma_0/2}\frac{\sin{(\gamma_0/2)}}{\gamma_0/2}\hat{S}_{j-1}^+\hat{S}_j^-\hat{S}_{j+1}^+\hat{S}_{j+2}^- \right. \nonumber\\
     &+& \left. e^{i\gamma_0}\frac{\sin{\gamma_0}}{\gamma_0}\hat{S}_{j-1}^+\hat{S}_j^+\hat{S}_{j+1}^-\hat{S}_{j+2}^- + \text{h.c.}\right],
     \label{eq:HF1aspinhalf}
\eea
where $\gamma_0=V_0T/\hbar$, see also~\cite{SI}. Table \ref{table:spin_model_F} also lists the frequencies $\omega^*$ at which the respective processes vanish in first-order as a result of DI. When $\omega^*\equiv\omega^*_e=V_0/(\hbar m)$, with $m$ being an integer, both the second and third terms in $\ham_1$ vanish, giving 
\beq
    \ham_F^{(1),e}=\frac{J}{2}\sum_{j=2}^{L-2} \left[\hat{S}_{j-1}^+\hat{S}_j^-\hat{S}_{j+1}^-\hat{S}_{j+2}^+ + \text{ h.c.}\right].
    \label{eq:HF1e}
\eeq
The only moves allowed by $\ham_F^{(1),e}$ are dipole-conserving ones $|+--+\rangle \leftrightarrow |-++-\rangle$, giving rise to Type-I constraint. In this notation, `$+$' refers to $S^z=1/2$ and `$-$' refers to $S^z=-1/2$. On the other hand, if $\omega^*\equiv\omega^*_o=2V_0/((2m+1)\hbar)$, only the third term is suppressed, giving
\bea
     \ham_F^{(1),o}&=&\frac{J}{2}\sum_{j=2}^{L-2} \Big[\hat{S}_{j-1}^+\hat{S}_j^-\hat{S}_{j+1}^-\hat{S}_{j+2}^+ \nonumber\\
     &+& \frac{i}{(m+\frac{1}{2})\pi} \hat{S}_{j-1}^+\hat{S}_j^-\hat{S}_{j+1}^+\hat{S}_{j+2}^- + \text{h.c.} \Big].
     \label{eq:HF1o}
\eea
In this case, the moves allowed are $|+--+\rangle \leftrightarrow |-++-\rangle$ and $|+-+-\rangle\leftrightarrow|-+-+\rangle$. This imposes constraint on the dynamics without conserving the dipole moment or any other obvious physical quantity, resulting in a Type-II constraint.

In \cite{SI}, we show that both $\ham_F^{(1),o}$ and $\ham_F^{(1),e}$ exhibit strong HSF and have an exponentially large number of frozen states in their spectrum. In the End Matter, we also show plots of the phase factors $f(t)$ to show that the corresponding trajectories interfere destructively over a full time period. We remark that a continuous drive protocol can also be used to obtain $\ham_F^{(1),o}$, but the same is not true for $\ham_F^{(1),e}$. The general discussion regarding the choice of drive protocol is relegated to the End Matter.

\begin{figure}
    \centering
    \includegraphics[width=1.0\linewidth]{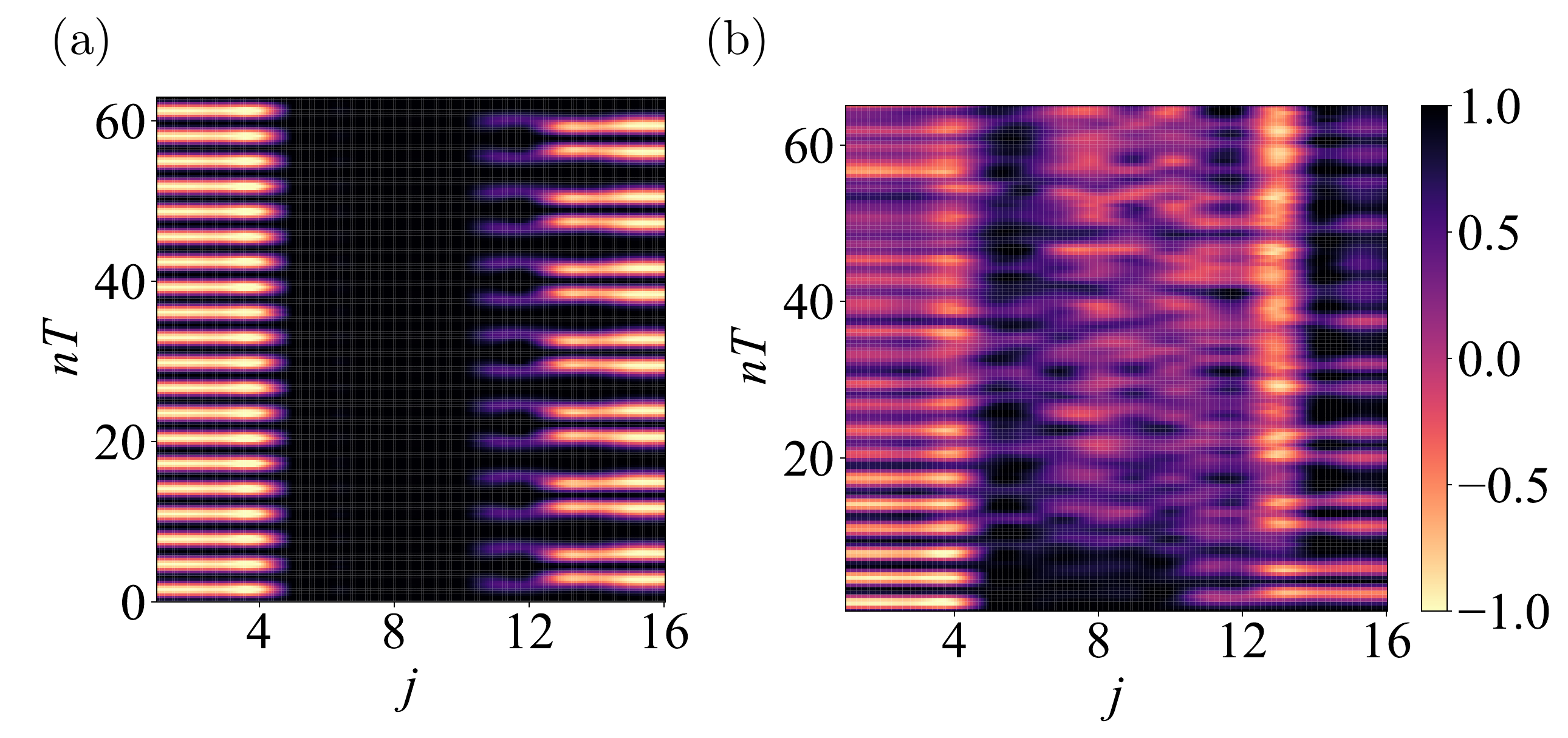}
    \caption{(Color online) Spatio-temporal profile of the OTOC $F(r,nT)$ starting from a state which, with respect to $\ham_F^{(1),e}$, has active blocks from sites $j=1-4$ and $j=11-16$, separated by an inactive block. The initial information in both the figures is localized at $i=2$ and $i=11$, i.e., $\sigma^z_i(0)\equiv \sigma_2^z\sigma_{11}^z$. (a) At $\omega^*_e=V_0/\hbar$, when the evolution is governed by $\ham_F^{(1),e}$ to leading order, information remains localized within the two active regions of space. (b) At $\omega_o^*=2V_0/(3\hbar)$, the two active regions communicate with each other within the first $100$ drive cycles, signifying that this state does not exhibit spatial localization under $\ham_F^{(1),o}$. All time evolutions are carried using exact Floquet operator $\uevol_F(T,0)$ and the drive amplitude $V_0=20$. See text for details.}
    \label{fig:fig3}
\end{figure}

{\it Spatial Localization --} We show next that both $\ham_F^{(1),e}$ and $\ham_F^{(1),o}$ allow disorder-free spatial localization in certain special kinds of initial states. The drive frequency thus acts like a tuning knob to steer in and out of localization. 

From the allowed processes mentioned in the last section, it is clear that a state having a domain wall structure like $|\ldots ++++----+++---\ldots\rangle$ is inert under both $\ham_F^{(1),o}$ and $\ham_F^{(1),e}$. Here, the ellipsis denote strings of up spins or down spins, with the separation between any two successive domain walls being at least 4 sites. If we now embed a motif like $+--+$ or $-++-$ (which can only move by dipole-conserving moves) within this sea of inert block, then such a motif cannot propagate through the chain. The reason is straightforward: a $+-$ dipole can only exchange position with a $-+$ dipole so that the constraint is obeyed. Whenever it encounters a wall of $++++$ or $----$ which is at least 3-site thick, it can no longer melt the wall and the propagation stops. Continuing this way, we can increase the size of the active region by embedding motifs which can only propagate by dipole-conserving moves and flanking them by inert blocks on either side. In this way, we can localize information within a limited region of space. For more details, we refer to \cite{SI}.

It is also possible to have states that exhibit localization when driven at $\omega^*_e$ but do not exhibit localization when driven at $\omega^*_o$. These states can be constructed by having motifs like $+-+-$ or $-+-+$ within the active block while satisfying all other criteria for embedding. This motif can melt the entire domain wall by virtue of the $\hat{S}_{j-1}^+\hat{S}_j^-\hat{S}_{j+1}^+\hat{S}_{j+2}^-$ term in $\ham_F^{(1),o}$, but is unable to do so in $\ham_F^{(1),e}$.  

The quantity that probes the scrambling of quantum information is the out-of-time-ordered correlator (OTOC), which in our case can be defined as $F(r,t=nT)=\langle\psi|\sigma_i^z(t)\sigma_j^z(0)\sigma_i^z(t)\sigma_j^z(0)|\psi\rangle$, where $\sigma_j^z=2S_j^z$ and $r=|j-i|$. This quantity is close to $1$ when information from site $i$ has not reached site $j$ in time $t$, whereas a lower value of $F(r,t)$ implies information scrambling~\cite{otoc1}.

We consider a state $|a\rangle=|+--+---+++-+-++- \rangle$ and show that it exhibits spatial localization when driven at $\omega^*_e$, but not at $\omega^*_o$. In Fig. \ref{fig:fig3}, we investigate the spatio-temporal profile of $F(r,nT)$ starting from the state $|a\rangle$ for a spin chain of length $L=16$ both at $\omega^*_e=V_0/\hbar$, see Fig.~\ref{fig:fig3}(a), and $\omega^*_o=2V_0/(3\hbar)$, see Fig.~\ref{fig:fig3}(b). Note that in these plots, the time evolution is carried out using the exact Floquet unitary, $\uevol_F(T,0)$. It can be seen that, with respect to $\ham_F^{(1),e}$, this state has two active blocks from sites $1-4$ and sites $11-16$ separated by an inactive block. In Figs.~\ref{fig:fig3}(a) and~\ref{fig:fig3}(b), we set the drive amplitude to $V_0=20$. The initial operator has weight on sites $2$ and $11$, i.e., $\sigma_i^z(0)\equiv \sigma_2^z\sigma_{11}^z$. In Fig. \ref{fig:fig3}(a), as time progresses, the operator $\sigma_2^z$ spreads in the region spanning the sites $j=1-4$ (evident from the value of $F(r,nT)$ reducing from 1 here), whereas the operator $\sigma_{11}^z$ spreads in the region $j=11-16$. Almost no information propagates through the inactive block ranging from $j=5-10$ \cite{note2}. Figure~\ref{fig:fig3}(b) shows the same spatio-temporal profile, but now at $\omega_o^*=2V_0/(3\hbar)$. The initial operator is the same as in the previous case. However, in this case, the information spreads throughout the entire spin chain within the first $n=100$ drive cycles. This is because of the $+-+-$ motif from sites $10-13$ which can propagate throughout the chain by displacing single $+$ and $-$ charges by 4 sites under the action of $\ham_F^{(1),o}$ (see also \cite{SI}).

In the Supplementary Material \cite{SI}, we provide example of states which exhibit localization both at $\omega^*_e$ and $\omega^*_o$, and we show plots of another quantity, namely the mutual information, which also acts as a diagnostic of the spread of quantum information.

{\it Discussion --} We propose a general unifying framework to generate kinetic constraints in Floquet systems by exploiting the principle of destructive many-body interference between quantum trajectories at special frequencies. %We identify special frequencies at which the quantum trajectories for processes violating the constraints interfere destructively over a complete drive cycle in first-order, approximately suppressing them. 
This interference occurs in first-order and given the complexity of the phenomenon, we expect that exact destructive interference will not occur in general, when higher-order corrections are taken into account. Nevertheless, the manifestation of these constraints will be observed in the dynamics of the system for an exponentially long prethermal timescale \cite{sghosh1,scaling1,scaling2,scaling3}.
We note that the mechanism of quantum interference has been explored in some other contexts to suppress quantum tunneling~\cite{di1,di2}, but not in the context of constraint generation and Hilbert space fragmentation, to the best of our knowledge.

As an example of this principle, we generate constraints in a 1-d spin-$1/2$ chain by periodically driving a dipole moment term in the Hamiltonian. Depending on the drive protocol and drive frequency, we show that destructive interference can be either `complete' or `partial', which ultimately leads to realization of constraints associated with or without a globally conserved quantity, respectively. In this particular case, both constraints give rise to strong HSF, which further leads to spatial localization of quantum information depending on the initial state. This provides the first example of HSF-induced spatial localization in the Floquet setting, thereby providing a frequency control to tune in and out of localization.

We finally outline that several other ergodicity-breaking mechanisms in driven systems, namely dynamical freezing due to emergent symmetries \cite{adas1,adas2,dsen1}, quantum many-body scars \cite{flscar1,flscar2,flscar5}, and previous realizations of HSF \cite{sghosh1,dsen2} can all be explained through the lens of destructive many-body interference. Moreover, this framework also has the potential of generating other constrained models in an out-of-equilibrium setting, including the Rydberg constraint. It remains to be tested whether this framework is able to generate quantum versions of kinetically constrained models inspired by classical glassy dynamics, including the number-conserving East and the North-East models \cite{gd1,gd2,gd3}. 

\textit{Acknowledgements--} S.G. and L.V. acknowledge support from the Slovenian Research and Innovation Agency (ARIS), Research core funding Grants No.~P1-0044, N1-0273 and J1-50005, as well as the Consolidator Grant Boundary-101126364 of the European Research Council (ERC).
S.G. and L.V. gratefully acknowledge the High Performance Computing Research Infrastructure Eastern Region (HCP RIVR) consortium~\cite{vega1} and European High Performance Computing Joint Undertaking (EuroHPC JU)~\cite{vega2}  for funding this research by providing computing resources of the
HPC system Vega at the Institute of Information sciences~\cite{vega3}. S.G. thanks Roopayan Ghosh and Riccardo Senese for related discussions. K.S. thanks DST, India for support through the SERB project JCB/2021/000030.

\section{End Matter}
{\it Appendix A: First-order Floquet Unitary --} For the time-periodic Hamiltonian in Eq. (\ref{eq:ham_gen}), the time-evolution operator over a complete time period $T$ can be split in two parts, namely $\uevol_F\equiv\uevol(T,0)=\uevol_0(T,0)W(T,0)$. Here $\uevol_0(T,0)$ encodes the time evolution generated by $\ham_0(t)$ and is straightforwardly given by $\uevol_0(t,0)=\exp{\left(-i\Phi(t,0)\hat{Q} \right)}$, where $\Phi(t,0)=\frac{1}{\hbar}\int_0^t dt' V(t')$. 

The second term captures the effect of $\ham_1$ on time evolution and can be shown to satisfy the differential equation $i\hbar\frac{\partial W}{\partial t}=\left(\uevol_0^\dagger (t,0)\ham_1 \uevol_0(t,0)  \right)W$. The perturbative series solution of this equation is discussed in details in \cite{SI}. For our purpose, we note that the first-order result is given by
$\uevol_F^{(1)}=-\frac{i}{\hbar}\uevol_0(T,0)\int_0 ^T dt \uevol_0^\dagger (t,0)\ham_1 \uevol_0(t,0)$. Thus, to first-order, the transition amplitude is
\bea
    \langle n| \uevol_F^{(1)}|m\rangle &=& -\frac{i}{\hbar} \int_0^T dt \exp{\left(-i\Phi(T,t)Q^{(n)}\right)}\nonumber\\
     &&\langle n|\ham_1|m\rangle \exp{\left(-i\Phi(t,0)Q^{(m)}\right)}. 
     \label{eq:trans_amp}
\eea
This integrand forms the mathematical basis for the phase factors in Eq. (\ref{eq:phaselag}) and Fig. \ref{fig:fig2}.

{\it Appendix B: Choice of drive protocol --} The framework of DI also dictates the kind of drive protocol that should be chosen to generate the constraint. This depends on the values of $\Delta Q_j$'s associated with the processes that need to be suppressed for the constrained dynamics. We denote this set by $\{\Delta Q_j^*\}$.

To understand this, note that for imposing the constraint, the renormalization factor $F$ in Eq. (\ref{eq:HF1_gen}) needs to have at least one common solution at $\omega_D=\omega^*$ satisfying $F\left(\frac{V_0\Delta Q^*_j}{\hbar\omega^*} \right)=0$ for all $\Delta Q_j^*$'s. That is, there must be at least one frequency at which all the trajectories corresponding to each $\Delta Q^*_j$ interfere destructively. Below, we consider the two most common cases for the values of $\Delta Q_j^*$, and for each case we suggest what kind of drive protocol might serve the purpose.
\begin{enumerate}
    \item \textbf{$\Delta Q_j^*$ has two non-zero values of the form $\pm \Delta q$}: In this case, the function $F$ must have at least one root and be an even function in frequency space. Both the square pulse drive protocol, $V(t)=V_0 \text{sgn}(\sin{\omega_D t})$, and the continuous drive protocol, $V(t)=V_0 \sin{\omega_D t}$, satisfy this criterion. 
    \item \textbf{$\Delta Q_j^*$ are equispaced, i.e., $\Delta Q_j^*= p \Delta Q_\text{min}$ where $p$ is an integer:} In this case, $F$ needs to have roots at periodic intervals $\omega_p^*=\omega^*/p$, where $\omega^*$ satisfies $F\left(\frac{V_0\Delta Q_{min}}{\hbar\omega^*}\right)=0$. The square pulse protocol meets this criterion, but the continuous drive protocol does not. 
\end{enumerate}
For obtaining $\ham_F^{(1),o}$ in our spin-$1/2$ example, we only need to suppress the process with $\Delta p =\pm 4$. This belongs to the first category, and hence, either of square pulse or continuous drive protocol does the job. However, for $\ham_F^{(1),e}$, we need to suppress processes with $\Delta p=\pm 2,\pm 4$. This belongs to the second category and hence can be achieved only using the square pulse drive protocol.

Finally, we remark that other than these, it could still be possible to design specific drive protocols that meet the above criteria for a given set of $\{\Delta Q^*_j\}$. However, in general, if there are multiple values of $\Delta Q_j^*$ which do not have a simple relationship between each other, it might be difficult or even impossible to design a protocol which vanishes at a single $\omega_D$ for all values of $\Delta Q_j^*$.

\begin{figure}
    \centering
    \includegraphics[width=1.0\linewidth]{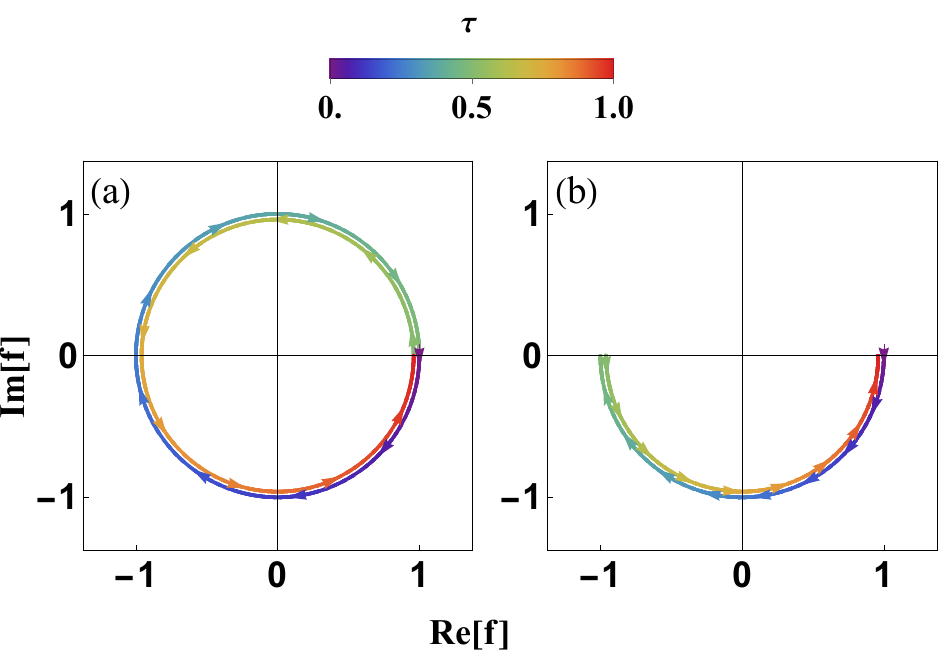}
    \caption{(Color online) Plot of the locus of $f(t)$ (Eq. (\ref{eq:phaselag}) and discussion before Eq. (\ref{eq:HF1aspinhalf})) as $t:0\to T$ for the process (a) $\hat{S}_{j-1}^+\hat{S}_j^+\hat{S}_{j+1}^-\hat{S}_{j+2}^-$ with $\Delta p=-4$ and (b) $\hat{S}_{j-1}^+\hat{S}_j^-\hat{S}_{j+1}^+\hat{S}_{j+2}^-$ with $\Delta p=-2$ in $\ham_1$. The renormalization factor $F$ for each process is obtained by summing $f(t)$ over a full period (Eq. (\ref{eq:UF1mn})). In (a), it is clear that $f(t)$ adds up to $F=0$, exhibiting DI and suppressing this process in first-order; whereas in (b), $f(t)$ has a resultant non-zero contribution when summed up. This results in the constrained Hamiltonian $\ham_F^{(1),o}$ in Eq. (\ref{eq:HF1o}). $\tau=t/T$.} 
    \label{fig:fig4}
\end{figure}

{\it Appendix C: Phase factor $f(t)$ for the spin-$1/2$ model at $\omega^*_o=2V_0/\hbar$ --}  We plot the profile of the phase lag $f(t)$, see discussion before Eq.~(\ref{eq:HF1aspinhalf}), accumulated as $t$ evolves from $0$ to $T$. We study the processes $\hat{S}_{j-1}^+\hat{S}_j^+\hat{S}_{j+1}^-\hat{S}_{j+2}^-$ with $\Delta p=-4$, see Fig.~\ref{fig:fig4}(a), and $\hat{S}_{j-1}^+\hat{S}_j^-\hat{S}_{j+1}^+\hat{S}_{j+2}^-$ with $\Delta p=-2$, see Fig.~\ref{fig:fig4}(b). The system is driven with a frequency $\omega_D=\omega^*_o=2V_0/\hbar$ and $t$ is scaled in terms of the drive period $T$ as $\tau=t/T$. The net amplitude, $F$, for a process is found by integrating over the locus of $f$ within a complete period, see Eq.~(\ref{eq:UF1mn}).

Figure \ref{fig:fig4}(a) shows that the locus of $f(t)$ at $\omega^*_o$ for $\Delta p=-4$ chalks out a full circle about the origin in the Argand plane as $t:0\to T$. The flow is in the clockwise direction for the first half-cycle, whereas it is in the anti-clockwise direction for the next half-cycle. We note that in this plot and every other plot in this figure, the clockwise flows overlap exactly with the anti-clockwise flows; the small shifts seen are for the aid to the eyes. From this plot, it is obvious that the phase factors $f(t)$ cancel each other exactly over a complete cycle ($F=0$), and hence these transitions are suppressed in $\ham_F^{(1),o}$, giving rise to the emergent constraint. Figure \ref{fig:fig4}(b) shows that for $\Delta p=-2$, the real part of $f(t)$ adds up to zero when summed over a full period, but the imaginary part survives, giving a net non-zero amplitude. The process $\hat{S}_{j-1}^+\hat{S}_j^-\hat{S}_{j+1}^+\hat{S}_{j+2}^-$ is thus allowed in $\ham_F^{(1),o}$.

A similar analysis (not shown) of the profile of $f(t)$ shows that at $\omega_D=\omega_e^*$, both of these processes are suppressed in $\ham_F^{(1),e}$.


\begin{thebibliography}{99}
\bibitem{DQevaluation} $\Delta \hat{Q}_i$ is defined such that $e^{\frac{i}{\hbar}\Delta\hat{Q}_i}\hat{P}_i=e^{\frac{i}{\hbar}\sum_j^{'}\hat{Q}_j} \hat{P}_i e^{-\frac{i}{\hbar}\sum_j^{'}\hat{Q}_j}$, in which the prime on the sum indicates that the sum is over those $j$'s, for which the support of $\hat{Q}_j$ overlaps with the support of $\hat{P}_i$. %Note that, in practice, it might not always be possible to find a closed-form expression for $\Delta \hat{Q}_i$. In such cases, we resort to evaluating matrix elements of the relevant operators between the initial state $|m\rangle$ and final state $|n\rangle$ in the computational basis, so that $\langle n|e^{\frac{i}{\hbar}\sum_j^{'}\hat{Q}_j} \hat{P}_i e^{-\frac{i}{\hbar}\sum_j^{'}\hat{Q}_j}|m\rangle = e^{\frac{i}{\hbar}(\Delta \hat{Q}_i)^{(nm)}}\langle n |\hat{P}_i|m\rangle$, where $(\Delta \hat{Q}_i)^{(nm)}=\sum_j^{'} (Q_j^{(n)}-Q_j^{(m)})$, since $\hat{Q}_j$ is diagonal in the computational basis.
\bibitem{note2} Careful observation suggests that there is very little information leak through the inactive block. This is due to the fact that higher-order corrections to the Floquet Hamiltonian lift this fragmented structure, thereby destroying information localization. However, the lowest-order correction to $\ham_F$ is of the order of $O(J^3/\omega^2)$ so that at high drive frequencies, such leakage terms are suppressed, leading to almost complete localization of information.
\end{thebibliography}
\end{document}

\title{Supplementary Material:\\ ``Destructive Interference induced constraints in Floquet systems"}
\author{Somsubhra Ghosh$^1$, Indranil Paul$^2$,
K. Sengupta$^3$, Lev Vidmar${^{1,4}}$}
\affiliation{$^1$Department of Theoretical Physics, J. Stefan Institute, SI-1000 Ljubljana, Slovenia. \\
$^2$Universit\'{e} Paris Cit\'{e}, CNRS, Laboratoire Mat\'{e}riaux
et Ph\'{e}nom\`{e}nes Quantiques, 75205 Paris, France.\\
$^3$School of Physical Sciences, Indian Association for the Cultivation of Science, Kolkata 700032, India.\\
$^4$Department of Physics, Faculty of Mathematics and Physics, University of Ljubljana, SI-1000 Ljubljana, Slovenia\looseness=-1}
\date{\today}

%\pacs{03.75.Lm, 05.30.Jp, 05.30.Rt}

\maketitle

\section{Floquet Perturbation Theory}
\label{fpt}
\subsection{Derivation of $\ham_F^{(1)}$ for the general model, Eq. (4) in main text}
\label{HF1general}
In this section, we outline the procedure for deriving the first-order Floquet Hamiltonian $\ham_F^{(1)}$ for both the general case (Eq. (4)) and the specific model which we consider in the main text (Eq. (6)).

We consider short-range, time-periodic Hamiltonians of the form 
\bea
\ham(t)&=&\ham_0(t)+\ham_1, \text{where}\nonumber\\ 
\ham_0(t)&=&V(t)\hat{Q}\equiv V(t)\sum_j \hat{Q}_j \text{ and } \ham_1=\sum_j J_j \hat{P}_j,
\label{eq:Htgeneral}
\eea
where $\hat{P}_j$ represents some dynamical term, like hopping or spin-flip, and $\hat{Q}_j$ is a few-body operator, with $[\hat{P}_i,\hat{Q}_j]\neq 0$ in general. The operator $\hat{Q}_j$ is assumed to be diagonal in the computational basis, which is most often the case in the study of constrained systems. Here, $j$ is a collective index that refers to the sites over which the corresponding operator, $\hat{P}_j$ or $\hat{Q}_j$, has support. The energy parameter $V(t)$ is driven periodically with the drive amplitude $V_0\gg J_j$, for all $j$ and a time period $T=2\pi/\omega_D$. This regime ensures that the Floquet Hamiltonian can be reliably computed perturbatively using FPT or RWA \cite{fl1,fl2,fl3}.

 The time-evolution operator for a complete drive cycle $\uevol(T,0)$ can be split into two parts,
 $\uevol(T,0)=\uevol_0(T,0)W(T,0)$, where 
 \beq
 \uevol_0(T,0)=\exp{\left(-\frac{i}{\hbar}\int_0^T dt \ham_0(t) \right)}
 \eeq
 encodes the time evolution due to $\ham_0$, and $W(T,0)$ captures the effect of $\ham_1$. Thus, $\uevol_0(t,0)$ satisfies $i\hbar \frac{\partial\uevol_0}{\partial t}=\ham_0 \uevol_0$. Using this and the fact that the full evolution operator should satisfy $i\hbar\frac{\partial \uevol}{\partial t}=\ham \uevol$, it can be shown that $W(t)$ satisfies \cite{sghosh2}
 \beq
 i\hbar\frac{\partial W}{\partial t}=\left(\uevol_0^{\dagger}(t,0) \ham_1 \uevol_0(t,0) \right)W=\ham_1^I(t)W.
 \label{eq:diffW}
 \eeq
For the rest of the analysis, we assume that $\uevol_0(T,0)=\mathds 1$, since this is the case for most of the commonly used drive protocols.

Equation (\ref{eq:diffW}) has a series solution in the form of Dyson series,
\beq
W(T,0)=1-\frac{i}{\hbar} \int_0^T dt \ham_1^I(t)+\frac{1}{2}\left(-\frac{i}{\hbar}\right)^2 \int_0^T dt_1 \int_0^{t_1}dt_2 \ham_1^I(t_1)\ham_1^I(t_2)+\ldots. 
\label{eq:WDyson}
\eeq
By the Floquet theorem, the one-period time-evolution operator is given by $\uevol_F\equiv \uevol(T,0)=e^{-\frac{i}{\hbar} \ham_F T}$, where $\ham_F$ is the exact Floquet Hamiltonian. Expanding $\ham_F$ in a perturbative series with $J_j/V_0$ as the perturbation parameter, $\ham_F=\ham_F^{(0)}+\ham_F^{(1)}+\ham_F^{(2)}+\ldots$, and comparing order-by-order with Eq. (\ref{eq:WDyson}) gives \cite{sghosh2}
\begin{align}
 \ham_F^{(0)} &= \frac{i \hbar}{T} \log \uevol_0(T, 0) = 0,
 \label{eq:HF0}\\
  \ham_F^{(1)} &=  \frac{1}{T} \int_0^T dt \ham_1^I(t),
  \label{eq:HF1}\\
   \ham_F^{(2)} 
   &= -\frac{i}{2 \hbar T} \int_0^T d t_1 \int_0^{t_1} dt_2 
   \left[ \ham_1^I(t_1) , \ham_1^I(t_2) \right],
   \label{eq:HF2}\\
   \ham_F^{(3)} 
   &= - \frac{1}{6 \hbar^2T} \int_0^T d t_1 \int_0^{t_1} d t_2 \int_0^{t_2} d t_3
   \left\{ \left[ \ham_1^I(t_1) , 
   \left[ \ham_1^I(t_2) , \ham_1^I(t_3) \right] \right] 
   + \left[ \left[\ham_1^I(t_1) ,  \ham_1^I(t_2)  \right], \ham_1^I(t_3)  \right] 
   \right\}.
   \label{eq:HF3}
\end{align}

In this case, $\uevol_0(t,0)=\exp{\left(-i\Phi(t,0) \hat{Q} \right)}$, where $\Phi(t,0)=\frac{1}{\hbar}\int_0^t dt' 
 V(t')$.  We assume $\Phi(T,0)=0$, so that the zeroth-order Floquet unitary operator, $\uevol_{0}(T,0)=\mathds 1$, and the zeroth-order Floquet Hamiltonian (Eq. (\ref{eq:HF0})) is $\ham_F^{(0)}=0$. Thus, the first-order Floquet Hamiltonian (Eq. (\ref{eq:HF1})) is given by
\bea
    \ham_F^{(1)}&=&\frac{1}{T}\int_0^T dt \hspace{2 pt} \uevol_0^\dagger (t,0) \left( \sum_j J_j\hat{P}_j \right)\uevol_0(t,0)\nonumber\\
    &=& \frac{1}{T}\sum_j J_j \int_0^T dt \exp{\left(i \Phi(t,0)\Delta \hat{Q}_j \right)}\hat{P}_j.         
 \label{eq:HF1generala}
\eea
This is Eq. (4) in the main text with the function $F\left(\frac{V_0 \Delta \hat{Q}_j}{\hbar\omega_D}\right)$ defined as $F\left(\frac{V_0 \Delta \hat{Q}_j}{\hbar\omega_D}\right)=\frac{1}{T}\int_0^T dt \exp{\left(i\Phi(t,0)\Delta \hat{Q}_j\right)}$. 

As outlined in the main text and also illustrated in the spin-$1/2$ model in the next subsection, the $\Delta \hat{Q}_j$ operator is such that 
\beq
e^{\frac{i}{\hbar}\Delta\hat{Q}_j}\hat{P}_j=e^{\frac{i}{\hbar}\sum_k^{'}\hat{Q}_k} \hat{P}_j e^{-\frac{i}{\hbar}\sum_k^{'}\hat{Q}_k}, 
\eeq
in which the prime on the sum indicates that the sum is over those $k$'s, for which the support of $\hat{Q}_k$ overlaps with the support of $\hat{P}_j$. For an operator which is diagonal in the computational basis, in most cases, it will be possible to figure out the operator $\Delta\hat{Q}_j$ using Baker-Campbell-Hausdorff expansion. In cases when this is too complicated, it can also be determined by calculating the matrix elements of $\ham_F^{(1)}$ in the computational basis, as is also illustrated for the spin-$1/2$ model.

\subsection{Derivation of $\ham_F^{(1)}$ for spin model, Eq. (6) in main text}
\label{HF1spin}

We consider a one-dimensional spin-$1/2$ chain with spin-spin interactions between four nearest-neighbor spins and a periodically driven dipole term, viz
\bea
    \ham(t)&=&\ham_0(t)+\ham_1, \text{ where }\nonumber\\
    \ham_0(t)&=&V(t)\sum_{j=1}^L j\hat{S}_j^z = V(t)\hat{p},\nonumber\\
    \ham_1&=& \frac{J}{2}\sum_{j=2}^{L-2}[\hat{S}_{j-1}^+\hat{S}_j^-\hat{S}_{j+1}^-\hat{S}_{j+2}^+ + \hat{S}_{j-1}^+\hat{S}_j^-\hat{S}_{j+1}^+\hat{S}_{j+2}^-
    + \hat{S}_{j-1}^+\hat{S}_j^+\hat{S}_{j+1}^-\hat{S}_{j+2}^- + \text{h.c.}] 
    \label{eq:spin_half_ham}
\eea
The drive amplitude $V_0$ is taken to be the largest energy scale in the problem such that $V_0\gg J$ and FPT can be used to find $\ham_F$ perturbatively.

The square pulse protocol used in the main text is given by
\bea
V(t)&=&-V_0; \quad \text{for }0\leq t\leq T/2\nonumber\\
&=& V_0 \quad\quad \text{for }T/2\leq t\leq T.
\label{eq:sqpulse}
\eea
In this case, $\uevol_0(t,0)$ for the spin Hamiltonian in Eq. (\ref{eq:spin_half_ham}) is given by
\bea
\uevol_0(t,0)&=&e^{\frac{i}{\hbar}V_0 t\sum_j j\hat{S}^z_j };\quad \quad\text{for }0\leq t\leq T/2\nonumber\\
&=&e^{\frac{i}{\hbar}V_0(T-t)\sum_j j\hat{S}^z_j };\quad \text{for }T/2\leq t\leq T.
\label{eq:U0}
\eea
Thus, from Eq. (\ref{eq:HF0}), it can be immediately seen that the zeroth-order Floquet Hamiltonian $\ham_F^{(0)}=0$. Next, we calculate the first-order Floquet Hamiltonian $\ham_F^{(1)}$ (Eq. (\ref{eq:HF1})). Due to symmetry of the drive protocol ($\uevol_0(t)=\uevol_0(T-t)$), it can be shown that the integral in Eq. (\ref{eq:HF1}) reduces to
\beq
    \ham_F^{(1)}=\frac{2}{T}\int_0^{T/2} dt \hspace{2 pt} \uevol_{0,<}^\dagger (t,0) \ham_1\uevol_{0,<}(t,0).
    \label{eq:HF1aspinhalf}
\eeq  
Here, $\uevol_{0,<}(t,0)$ refers to the time evolution operator for times $t\leq T/2$, as in the first equation of Eq. (\ref{eq:U0}). We evaluate the integrand in Eq. (\ref{eq:HF1aspinhalf}) by two equivalent approaches - one by using the Baker-Campbell-Hausdorff expansion and the other by examining the action of $\ham_1$ on the basis states.

Let us consider one spin-flip term from $\ham_{1}$, say, $S_{j-1}^+S_j^+S_{j+1}^-S_{j+2}^-$ and examine how to evaluate the operator integrand. Since the driven term is an on-site potential and the spins on different sites commute with each other, the exponential of the sum in $\uevol_0(t,0)$ can be cast as a product of exponentials, i.e., $\uevol_0(t,0)=\prod_k e^{\frac{i}{\hbar}V_0t kS_k^z}$. Again, due to commutativity of spins at different sites, the only non-trivial terms corresponding to $S_{j-1}^+S_j^+S_{j+1}^-S_{j+2}^-$ in the integrand of Eq. (\ref{eq:HF1aspinhalf}) can be written as
\beq
    \uevol_{0,<}^\dagger (t,0) S_{j-1}^+S_j^+S_{j+1}^-S_{j+2}^-\uevol_{0,<}(t,0)= \prod_{k=j-1}^{j+2} e^{-\frac{i}{\hbar}V_0t kS_k^z} S_k^{a_k} e^{\frac{i}{\hbar}V_0t kS_k^z},
\eeq
where $a_{j-1},a_j=+$ and $a_{j+1},a_{j+2}=-$. Each of these terms can be evaluated using the Baker-Campbell-Hausdorff formula to give
\beq 
 e^{-\frac{i}{\hbar}V_0t kS_k^z} S_k^{a_k} e^{\frac{i}{\hbar}V_0t kS_k^z} =  e^{-\frac{2i}{\hbar}V_0t kS_k^z} S_k^{a_k}
 \eeq
 effectively. Thus,
\beq
    \uevol_{0,<}^\dagger (t,0) S_{j-1}^+S_j^+S_{j+1}^-S_{j+2}^-\uevol_{0,<}(t,0)=e^{-\frac{2i}{\hbar}V_0t\sum_{k=j-1}^{j+2} kS_k^z} S_{j-1}^+S_j^+S_{j+1}^-S_{j+2}^-.
    \label{eq:HF1bspinhalf}
\eeq
It can be checked that this holds true for every other term in $\ham_{1}$, so that
\beq
    \uevol_{0,<}^\dagger (t,0) \ham_{1}\uevol_{0,<}(t,0) = e^{-\frac{2i}{\hbar}V_0t\sum_{k=j-1}^{j+2} kS_k^z} S_{j-1}^{a_{j-1}}S_j^{a_j}S_{j+1}^{a_{j+1}}S_{j+2}^{a_{j+2}},
    \label{eq:HF1cspinhalf}
\eeq
where two of $a_k$'s are plus and the other two minus. 

Alternatively, the integrand can be simplified by examining its action on the computational basis states. We tabulate the action of the different spin-flip terms in $\ham_1$ on the basis states $|m\rangle$ below and for each of them, calculate the change in the total dipole moment.
\begin{center}
\begin{tabular}{ |c|c|c| } 
 \hline
 Term & $|m\rangle \quad \rightarrow \quad |n\rangle$ & $\Delta \left(\sum_k kS_k^z\right)$ \\ 
 \hline
 $S_{j-1}^+S_j^+S_{j+1}^-S_{j+2}^-$ & $|--++\rangle\to |++--\rangle$ & $-4 $ \\
 $S_{j-1}^+S_j^-S_{j+1}^+S_{j+2}^-$ & $|-+-+\rangle\to |+-+-\rangle$ & $-2$ \\
 $S_{j-1}^+S_j^-S_{j+1}^-S_{j+2}^+$ & $|-++-\rangle\to |+--+\rangle$ & $0$ \\
 \hline
\end{tabular}
\end{center}
Using this transition table, one obtains the same operator integrand as in Eq. (\ref{eq:HF1cspinhalf}). The integral in Eq. (\ref{eq:HF1aspinhalf}) can be computed as
\bea
    \ham_F^{(1)}=\frac{J}{2}\sum_{j=2}^{L-2} \hat{S}_{j-1}^+\hat{S}_j^-\hat{S}_{j+1}^-\hat{S}_{j+2}^+
    + e^{i\gamma_0/2}\frac{\sin{(\gamma_0/2)}}{\gamma_0/2}\hat{S}_{j-1}^+\hat{S}_j^-\hat{S}_{j+1}^+\hat{S}_{j+2}^-
    + e^{i\gamma_0}\frac{\sin{\gamma_0}}{\gamma_0}\hat{S}_{j-1}^+\hat{S}_j^+\hat{S}_{j+1}^-\hat{S}_{j+2}^- + \text{h.c.},
    \label{eq:HF1dspinhalf}
\eea
where $\gamma_0=V_0T/\hbar$. This gives Eq. (6) in the main text. At $\gamma_0=(2m+1)\pi$ (with $m$ being an integer), the last term in $\ham_F^{(1)}$ vanishes giving the fragmented Hamiltonian in Eq. (8) of the main text, viz
\bea
     \ham_F^{(1),o}=\frac{J}{2}\sum_{j=2}^{L-2} \Big(\hat{S}_{j-1}^+\hat{S}_j^-\hat{S}_{j+1}^-\hat{S}_{j+2}^+ 
     + \frac{i}{(m+\frac{1}{2})\pi} \hat{S}_{j-1}^+\hat{S}_j^-\hat{S}_{j+1}^+\hat{S}_{j+2}^- + \text{h.c.} \Big).
     \label{eq:HF1o}
\eea
At $\gamma_0=2m\pi$, both the second and third term vanishes, giving
\beq
    \ham_F^{(1),e}=\frac{J}{2}\sum_{j=2}^{L-2} \hat{S}_{j-1}^+\hat{S}_j^-\hat{S}_{j+1}^-\hat{S}_{j+2}^+ + \text{ h.c. }
    \label{eq:fragHF1b}
\eeq
 Both $\ham_F^{(1),o}$ and $\ham_F^{(1),e}$ admit of constrained dynamics, with the former not having an associated dipole moment conservation, while the dipole moment is conserved in the latter case.
 
\begin{figure}
     \centering
     \includegraphics[width=1.0\linewidth]{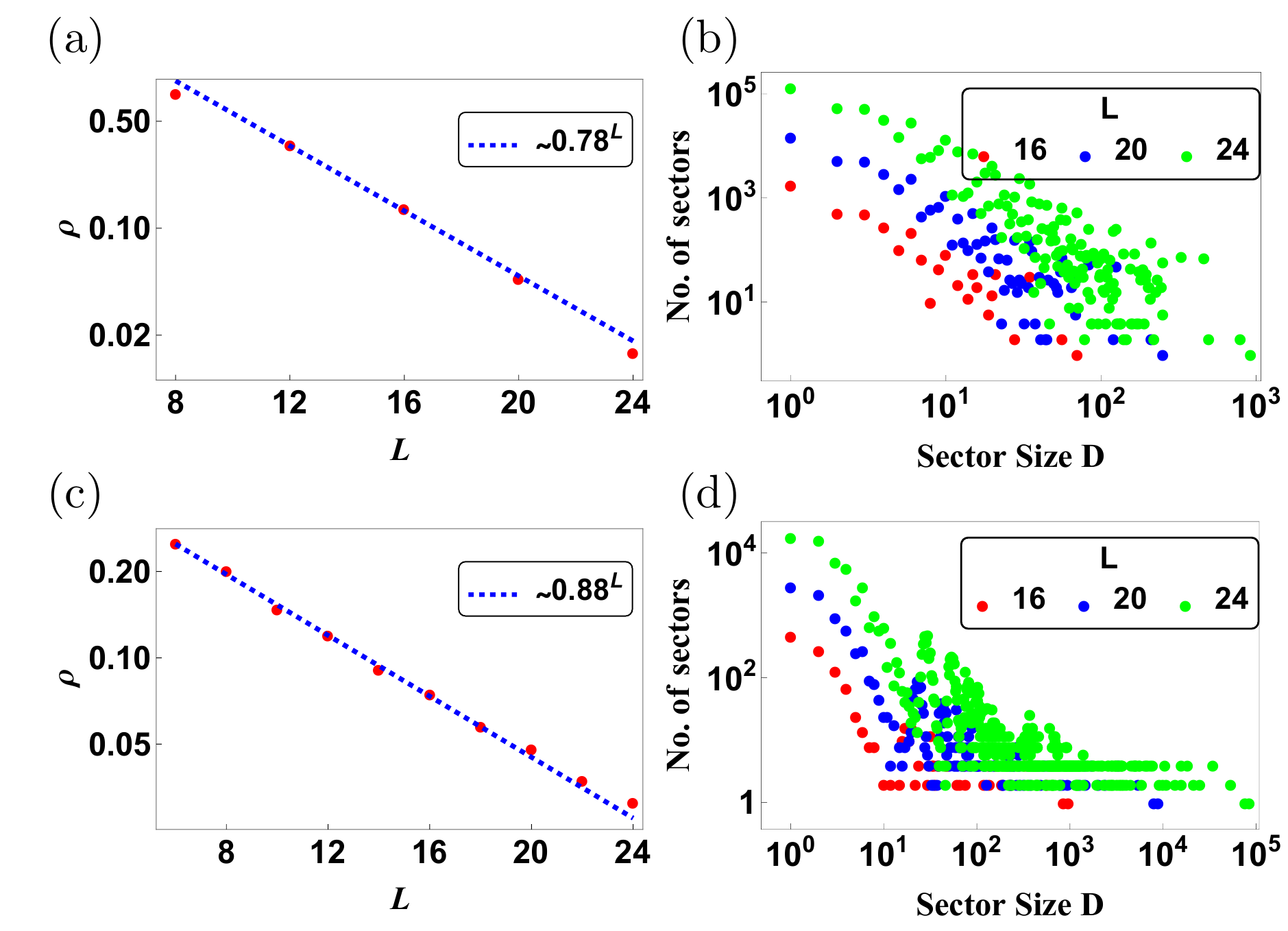}
     \caption{(Color online) (a) Plot of $\rho$ vs $L$ for the first-order Floquet Hamiltonian at $\gamma_0=2m\pi$, $\ham_F^{(1),e}$ (Eq. (\ref{eq:fragHF1b})) shows exponential decay with system size, indicating strong HSF. The largest symmetry sector corresponds to $S^z=0, p=0$ and the numerical fit (in blue dashed line) gives a scaling $\rho\sim (0.78)^L$. (b) Fragment size distribution for three different sizes $L=16$ (red), $20$ (blue), $24$ (green). This also shows initial algebraic decay followed by saturation for very large fragment sizes. (c) and (d) Similar analysis for the first-order Floquet Hamiltonian at $\gamma_0=(2m+1)\pi$, $\ham_F^{(1),o}$ (Eq. (\ref{eq:HF1o})) show that $\ham_F^{(1),o}$ also exhibits strong HSF. In this case, the largest symmetry sector is $S^z=0$. See text for details.}
     \label{fig:sfig1}
\end{figure}

 We plot the variation of the ratio $\rho=\mathcal{D}_\text{max}^\text{f}/\mathcal{D}_\text{max}^\text{s}$ versus system size $L$ and the sector size distribution for $\ham_F^{1,(e)}$ in Figs. \ref{fig:sfig1}(a) and \ref{fig:sfig1}(b) respectively; and the same quantities for $\ham_F^{(1),o}$ are probed in Figs. \ref{fig:sfig1}(c) and \ref{fig:sfig1}(d) respectively. Here, $\mathcal{D}_\text{max}^\text{f}$ refers to the dimension of the largest fragment and $\mathcal{D}_\text{max}^\text{s}$ refers to the dimension of the largest symmetry sector. Note that since the dipole moment is conserved for $\ham_F^{(1),e}$, we choose the sector $S^z=0$, $p=0$ as our largest symmetry sector in the plot for $\ham_F^{(1),e}$, but we choose the sector $S^z=0$ as the largest symmetry sector in case of $\ham_F^{(1),o}$. The exponential decay of $\rho$ with $L$ establishes that both $\ham_F^{(1),e}$ and $\ham_F^{(1),o}$ are strongly fragmented. We note, in passing, that these results hold true even if the drive protocol reads $V_1+V(t)\sum_{j=1}^L j\hat{S}_j^z$ with $V_1\ll V_0$.

\section{Enumeration of frozen states of $\ham_F^{(1),e}$ and $\ham_F^{(1),o}$}
\label{SI:frozen}

In this section, we use transfer matrices to establish that the number of frozen states (one-dimensional fragments) of $\ham_F^{(1),o}$ and $\ham_F^{(1),e}$ grows exponentially with the system size.

We consider $\ham_F^{(1),o}$ and focus on constructing frozen states iteratively starting with $L=4$. For $L=4$, the frozen states corresponding to this Hamiltonian can be written as
\bea
    && ++++, +++-, ++-+, +-++, -+++, ++--, \nonumber\\
    && --++, +---, -+--, --+-, ---+,---- .\nonumber
\eea
In total, there are $12$ frozen states for $L=4$, viz $N_f^{L=4}=12$.

We note here that the Hamiltonian $\ham_F^{(1),o}$ has four site interactions. Hence, a frozen state of size $L$ can become active in a chain of length $L+1$ only if the new $L+1$-th site forms an active block with the last 3 sites of the former. Thus, to construct frozen states of larger system sizes iteratively, we need to concentrate on the spin configurations on the last $3$ sites of a frozen state of the previous system size $L$ and check what possible configurations of the new spin on the $L+1$-th site will give a frozen state of size $L+1$. For instance, if a frozen state of length $L$ ends in $++-$, then adding a $+$ or a $-$ at the $L+1$-th site will give a frozen state of length $L+1$, ending in $+-+$ or $+--$, respectively. On the other hand, if a frozen state of length $L$ ends in $+-+$, then only a $+$ on the $L+1$-th site will give a frozen state of length $L+1$.

The following table enumerates all such possibilities. The left column gives the configurations of the last $3$ sites of a frozen state of length $L$, whereas the right column gives the possible options for the $L+1$-th site to generate a frozen state of size $L+1$.
\begin{center}
\begin{tabular}{ |c|c| } 
 \hline
  Last 3 sites of a frozen state of length $L$ & Options for $L+1$-th site to construct \\ & a frozen state \\ 
 \hline
 $+++$ & $+,-$ \\
 $---$ & $+,-$  \\
 $--+$ & $+,-$  \\
 $++-$ & $+,-$ \\
 $-+-$ & $-$ \\
 $+--$ & $-$ \\
 $-++$ & $+$ \\
 $+-+$ & $+$\\
 \hline
\end{tabular}
\end{center}
Using this table, it is possible to find out the number of frozen states of size $L+1$ from the number of frozen states of size $L$. The transfer matrix for the process can be written as 
\[
\begin{pmatrix}
N_{+++}^{L+1} \\
N_{---}^{L+1}\\
N_{--+}^{L+1}\\
N_{++-}^{L+1}\\
N_{-+-}^{L+1}\\
N_{+--}^{L+1}\\
N_{-++}^{L+1}\\
N_{+-+}^{L+1}
\end{pmatrix}
=
\begin{pmatrix}
1 & 0 & 0 &0&0&0&1&0 \\
0&1&0&0&0&1&0&0\\
0&1&0&0&0&0&0&0\\
1&0&0&0&0&0&0&0\\
0&0&1&0&0&0&0&0\\
0&0&0&1&1&0&0&0\\
0&0&1&0&0&0&0&1\\
0&0&0&1&0&0&0&0
\end{pmatrix}
\begin{pmatrix}
N_{+++}^{L} \\
N_{---}^{L}\\
N_{--+}^{L}\\
N_{++-}^{L}\\
N_{-+-}^{L}\\
N_{+--}^{L}\\
N_{-++}^{L}\\
N_{+-+}^{L}
\end{pmatrix},
\]
where $N^L_{abc}$ is the number of frozen states of length $L$ ending in the spin configuration $abc$. Summing all these $N^L_{abc}$'s will give the total number of frozen states of size $L$.

\begin{figure}
     \centering
     \includegraphics[width=1.0\linewidth]{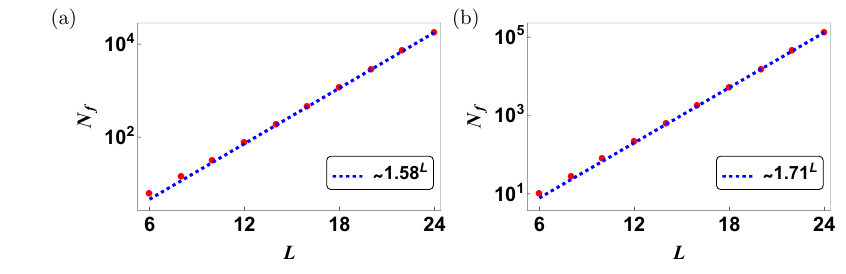}
     \caption{(Color online) Exponential scaling of number of frozen states with system size $L$ for (a) $\ham_F^{(1),o}$ and (b) $\ham_F^{(1),e}$. The numerically fitted lines are plotted using blue, dashed lines. See text for details.}
     \label{fig:sfig2}
\end{figure}

This can be solved exactly, in principle, by knowing the number of frozen states for $L=4$ and then going upward iteratively. But asymptotically, for large $L$, the scaling of this number is controlled by the largest eigenvalue $\lambda_{\rm max}$ of the transfer matrix. In this case, $\lambda_{\rm max} \sim 1.61$. Thus the number of frozen states for size $L$ scales asymptotically as $N_f\sim (1.61)^L$. We plot the numerical result for the scaling of $N_f$ with $L$ in Fig. \ref{fig:sfig2}(a). The scaling obtained from a least-square fit is $\sim (1.58)^L$, which is close to our analytical prediction.

A similar analysis can be done for the scaling of the number of frozen states of $\ham_F^{(1),e}$. The transfer matrix enumerating the number of frozen states can be written down following the same logic as in the previous case. Only two of the entries, which were previously zero, become non-zero now, resulting in
\[
\begin{pmatrix}
N_{+++}^{L+1} \\
N_{---}^{L+1}\\
N_{--+}^{L+1}\\
N_{++-}^{L+1}\\
N_{-+-}^{L+1}\\
N_{+--}^{L+1}\\
N_{-++}^{L+1}\\
N_{+-+}^{L+1}
\end{pmatrix}
=
\begin{pmatrix}
1 & 0 & 0 &0&0&0&1&0 \\
0&1&0&0&0&1&0&0\\
0&1&0&0&0&0&0&0\\
1&0&0&0&0&0&0&0\\
0&0&1&0&0&0&0&1\\
0&0&0&1&1&0&0&0\\
0&0&1&0&0&0&0&1\\
0&0&0&1&1&0&0&0
\end{pmatrix}
\begin{pmatrix}
N_{+++}^{L} \\
N_{---}^{L}\\
N_{--+}^{L}\\
N_{++-}^{L}\\
N_{-+-}^{L}\\
N_{+--}^{L}\\
N_{-++}^{L}\\
N_{+-+}^{L}
\end{pmatrix}.
\]
The largest eigenvalue of this matrix is $\lambda_{\rm max}\sim1.75$. So the asymptotic scaling of the number of frozen states is $N_f\sim (1.75)^L$. We plot the numerical result and the least-square fit in Fig. \ref{fig:sfig2}(b), where it can be seen that $N_f\sim (1.71)^L$. This is again close to our analytical result.

Physically, in both the cases, this exponential scaling stems from the fact that it is possible to construct more than one frozen state of size $L+1$ from one frozen state of size $L$ (as is evident from the transfer matrices). This effect continues to proliferate, giving the exponential scaling with $L$.

\section{The renormalization factor $F$ for commonly used drive protocols}
\label{SI:protocol}
In this section, we explicitly list the renormalization factors $F\left(\frac{V_0\Delta Q_j}{\hbar\omega_D} \right)$, (see text after Eq. (3) in main text) corresponding to the three most commonly used drive protocols. Depending on the processes $\hat{P}_j$'s that need to be suppressed and their associated $\Delta Q_j$'s, we discuss which protocol will be suitable for obtaining the constraints for a given case. See also Appendix B in End Matter of main text.

\subsection{Square Pulse}
The square pulse protocol is given by 
\bea
V(t)&=&-V_0; \quad \text{for }0\leq t\leq T/2\nonumber\\
&=& V_0 \quad\quad \text{for }T/2\leq t\leq T.
\label{eq:protocol_sqpulse}
\eea
Using the definition of $\Phi(t,0)$ and the phase factor $f(t)$ from Eq. (2) in the main text, $f(t)$ for the square pulse is given by
\beq
f(t)=\Theta(T/2-t)e^{iV_0t\Delta Q_j/\hbar}+\Theta(t-T/2)e^{iV_0(T-t)\Delta Q_j/\hbar},
\label{eq:ftsquare}
\eeq
where $\Theta$ is the Heaviside step function.
The total phase lag accumulated over a complete period $t$, which is also the renormalization factor, is obtained by integrating this $f(t)$ from $0$ to $T$, thus giving
\beq
F=\exp{\left(i\gamma_0\right)}\frac{\sin{\gamma_0}}{\gamma_0}
\label{eq:Fsquare},
\eeq
where $\gamma_0=\frac{\pi V_0\Delta Q_j}{2\hbar\omega_D}$, with $T=2\pi/\omega_D$. In this case $F$ has periodic roots given by $\gamma_0^*=m\pi$, with $m$ being an integer. Thus, the square pulse protocol is suitable in both cases: when the constraint involves suppression of processes with $\Delta Q_j^*=\pm \Delta q$, or the constraint involves suppression of processes with $\Delta Q_j^*=p\Delta Q_\text{min}$ ($p$ being an integer).

\subsection{Continuous drive protocol}
The continuous cosine drive protocol is given by $V(t)=V_0\cos{\omega_D t}$. The phase factor $f(t)$ is given by $f(t)=\exp{\left(i\frac{V_0\Delta Q_j}{\hbar\omega_D}\sin{\omega_Dt} \right)}$. Using the expansion $e^{i x \sin{\omega_Dt}}=\sum_{m=-\infty}^{\infty} \mathcal{J}_m(x)e^{im\omega_Dt}$, where $\mathcal{J}_m$ is the $m$-th order Bessel function of the first kind, we get for the renormalization factor
\beq
F=\mathcal{J}_0\left( \frac{V_0\Delta Q_j}{\hbar\omega_D}\right).
\label{eq:Fcosine}
\eeq
The zeroth order Bessel function does not admit of periodic roots but nevertheless, is a periodic function of its argument. Thus the cosine drive protocol is suitable for the case when the constraint requires suppression of processes with $\Delta Q_j^*=\pm q$, but not for the case which involves suppression of processes with $\Delta Q_j^*=p \Delta Q_\text{min}$.

\subsection{Kick protocol}
We finally consider the kick protocol $V(t)=V_0\delta (t-t_0-nT)$, where $0\leq t_0\leq T$. Note that we allow the kick to be anywhere within a period, not necessarily at one end of the period, as is commonly considered. There is a major difference here from the previous two cases. The drive protocol over a full period does not sum up to zero, so that $\uevol_0(T,0)\neq \mathds 1$. Instead, following the notation of the main text, $\Phi(T,0)=V_0/\hbar$. This has the consequence that $\ham_F^{(1)}$ cannot be directly written as Eq. (6); nevertheless, the first-order transition amplitude is still proportional to this term \cite{roopayan}. 

In this case, $f(t)=e^{-\frac{i}{\hbar} V_0 Q}\left[\Theta(t_0-t)+\Theta(t-t_0)e^{\frac{i}{\hbar}V_0\Delta Q_j}\right]$. Integrating over a full period, this gives for the renormalization factor
\beq
F=e^{-\frac{i}{\hbar} V_0 Q}\left[t_0+(T-t_0)\exp{\left(\frac{i}{\hbar}V_0\Delta Q_j \right)}\right].
\label{eq:Fdelta}
\eeq
The vanishing of this function requires $t_0=T/2$ and $V_0\Delta Q_j=(2m+1)\pi\hbar$, with $m$ being an integer. 

Thus, a kick protocol with the kick instance being at $T/2$ is suitable to generate constraints in cases when the constraint involves suppression of processes with $\Delta Q_j^*=\pm \Delta q$ or suppression of processes with $\Delta Q_j^*=(2p+1)\Delta Q_\text{min}$, with $p$ being an integer and $\Delta Q_\text{min}$ satisfying $V_0\Delta Q_\text{min}=(2m+1)\pi\hbar$.

Note that, in general, the phase factor $f(t)$ is a well-behaved function satisfying Dirichlet's condition. Thus $f(t)$ can be expanded in a Fourier series within the time slice $0$ to $T$, viz $f(t)=\sum_n g_n (\omega_D) e^{in\omega_D t}$, where $g_n$ is the $n$-th Fourier component. By the definition of the renormalization factor $F$, it is easily seen that $F\equiv g_0$. Thus, the properties which we ask of $F$ are actually properties of $g_0$. One thing to note here is that the mathematical form of the function $g_0$ is dictated by the chosen drive protocol whereas its value depends on $\omega_D,V_0$ and $\Delta Q_j$'s. This link paves the way for the design of other drive protocols which have a common root $\omega_D$ for all the given $\Delta Q_j^*$'s, which need to be suppressed.

\section{Spatial Localization}
\label{localization}
\subsection{Inactive and Active Blocks}
\label{blocks}

We use this section to expand on the mechanism of spatial localization in our spin model, which was discussed briefly in the main text. 

At the special frequencies given by $\gamma_0=(2m+1)\pi$, the only dynamics allowed by $\ham_F^{(1),o}$ (Eq. (8) in main text) are $|\ldots+--+\ldots\rangle \leftrightarrow |\ldots-++-\ldots\rangle$ and $|\ldots+-+-\ldots\rangle\leftrightarrow |\ldots-+-+\ldots\rangle$. The first is a dipole moment-conserving move where a $+-$ dipole can exchange position with an oppositely aligned dipole, viz $-+$ and vice versa. This ensures that the dipole moment is conserved. The second is equivalent to the shift of a single charge by 4 sites, so that the dipole moment changes by $\pm 2$. This can be understood from the following table. 
\begin{center}
\begin{tabular}{ |c| } 
\hline
 $\big|\mathcircled{+}\boxed{+-+-}\ldots\big\rangle \to \big|\boxed{+-+-}\mathcircled{+}\ldots\big\rangle$ \\
 \hline
 $\big|\boxed{-+-+}\mathcircled{-}\ldots\big\rangle \to \big|\mathcircled{-}\boxed{-+-+}\ldots\big\rangle$ \\
 \hline
 $\big|\mathcircled{-}\boxed{-+-+}\ldots\big\rangle \to \big|\boxed{-+-+}\mathcircled{-}\ldots\big\rangle$ \\
 \hline
 $\big|\mathcircled{+}\boxed{-+-+}\ldots\big\rangle \to \big|\boxed{-+-+}\mathcircled{+}\ldots\big\rangle$ \\
 \hline
 $\big|\boxed{+-+-}\mathcircled{-}\ldots\big\rangle \to \big|\mathcircled{-}\boxed{+-+-}\ldots\big\rangle$ \\
 \hline
 $\big|\boxed{-+-+}\mathcircled{+}\ldots\big\rangle \to \big|\mathcircled{+}\boxed{-+-+}\ldots\big\rangle$ \\
 \hline
\end{tabular}
\end{center}
In each case, the dynamics can be interpreted as a single charge (circled) passing through either a $+-+-$ or $-+-+$ motif. Consider, for example, the first move: The $+-+-$ motif extending from the second to the fifth site flips to $-+-+$ motif under $\ham_F^{(1),o}$. But this can be equivalently interpreted as the motion of the circled positive charge from the first site to the fifth site.

A domain wall state of the form $|\ldots ++++----+++---\ldots\rangle$, where $\ldots$ represent strings of up- or down- spins with the separation between any two successive domain being at least three sites, is inactive under this set of dynamics. This can melt neither by dipole moment-conserving moves nor by shifting isolated charges by four sites. Thus, this forms an inert or frozen state under $\ham_F^{(1),o}$. The presence of such inert structures allows us to localize quantum information by embedding active regions within inactive blocks. Note that when we refer to embedding active blocks here, we are referring to active blocks which can move by dipole moment-conserving moves only. 

The simplest example of such a state is having a motif of $+--+$ or a motif of $-++-$ embedded within a sea of inert blocks of $++++$ and $----$ on either side. A dipole of `$+-$' type moves by swapping positions with a dipole of `$-+$' type and vice versa. As soon as it encounters a wall of up spins or down spins, the propagation stops. This same strategy also facilitates to spatially separate different active blocks within the chain, by inserting inactive blocks, which are at least three sites thick, between them. The following shows some examples of this embedding. The active blocks are marked in boxes. 
\begin{center}
\bea
 &|++++\boxed{+--+}++++\rangle&\leftrightarrow |++++\boxed{-++-}++++\rangle\nonumber \\
 &|----\boxed{+--+}++++\rangle&\leftrightarrow |----\boxed{-++-}++++\rangle\nonumber  \\
 &|++\boxed{+--+-++-}++\rangle&\leftrightarrow
 \begin{tikzpicture}[baseline]
    \node[anchor=base] at (0,0) {$|++\boxed{-++--++-}++\rangle$};
    \draw[<->, thick] (1.5,0.5) -- (2,1);  % Diagonal up arrow
    \draw[<->, thick] (1.5,-0.5) -- (2,-1); % Diagonal down arrow
    \node[anchor=base] at (3,1.5) {$|++\boxed{-+-++-+-}++\rangle$};
    \node[anchor=base] at (3,-1.5) {$|++\boxed{-++-+--+}++\rangle$};
  \end{tikzpicture}
  \label{eq:actinact}
\eea
\end{center}
This immediately suggests that with $\ham_F^{(1),o}$ alone, it might be possible to spatially localize information in this spin chain. Higher-order terms in the Floquet Hamiltonian will tend to destroy this localization and scramble the information throughout the chain, but at high drive amplitudes (and hence high drive frequencies), we expect this delocalization to set in after an exponentially long prethermal timescale.

We discussed the case of $\ham_F^{(1),o}$ here. The discussion for $\ham_F^{(1),e}$ will involve the dipole-conserving term only and would be similar.

\subsection{Out-of-Time Ordered Correlator (OTOC)}
\label{otoc}

We have discussed in the main text that the OTOC is a quantity which can probe the spread of quantum information in a quantum system. We use this section to elaborate on this and show schematically how OTOC can detect spatial localization in our model.

For our case, we define the OTOC as $F(r,t=nT)=\langle\psi|\sigma_i^z(t)\sigma_j^z(0)\sigma_i^z(t)\sigma_j^z(0)|\psi\rangle$, where $\sigma_j^z=2S_j^z$ and $r=|j-i|$. This quantity essentially investigates how an information initially localized at site $i$ spreads to site $j$ in time $t$, and can be shown to be related to the squared commutator $C(r,t=nT)= \langle\psi|[\sigma_i^z(t),\sigma_j^z(0)]^\dagger[\sigma_i^z(t),\sigma_j^z(0)]|\psi\rangle$ as $C(r,t)=2(1-F(r,t))$. Thus, a higher value of $F(r,t)$ (close to 1) would imply that the information from site $i$ has not reached $j$ in time $t$, whereas a lower value of $F(r,t)$ would imply information scrambling. 

Let us start from a Fock state having an active block embedded within inert regions, viz $|\psi\rangle=|\begin{tikzpicture}
    % Draw the oval next to the rectangle
    \draw[fill=red!25] (0.1, 0.15) ellipse (0.5 and 0.25);
    % Draw the rectangle
    \draw[fill=blue!25] (0.6, -0.1) rectangle (1.6, 0.4);
    \draw[fill=red!25] (2.1, 0.15) ellipse (0.5 and 0.25);
\end{tikzpicture}\rangle$, where the center rectangular region is active and the two ellipses on the sides are inactive. We consider time evolution with $\ham_F^{(1)}$ and choose the initial site $i$ to be within the active region and the probe site $j$ to be within the inactive block. As discussed and illustrated above, the inactive blocks remain unaffected by the action of $\ham_F^{(1)}$, it is only the configuration of the active block which changes. Due to this fact, the state $|\psi\rangle$ is an eigenstate for both the $\sigma_j^z(0)$ operators which appear in the expression of $F(r,t)$. This can be seen more clearly as follows
\bea
    & &\sigma_i^z(t)\sigma_j^z(0)\sigma_i^z(t)\sigma_j^z(0)|\psi\rangle\nonumber\\
    & =&\sigma_i^z(t)\sigma_j^z(0)\sigma_i^z(t)\sigma_j^z(0)\quad|\begin{tikzpicture}
    % Draw the oval next to the rectangle
    \draw[fill=red!25] (0.1, 0.15) ellipse (0.5 and 0.15);
    % Draw the rectangle
    \draw[fill=blue!25] (0.6, -0.1) rectangle (1.6, 0.4);
    \draw[fill=red!25] (2.1, 0.15) ellipse (0.5 and 0.15);
    \end{tikzpicture}\rangle\nonumber\\
    &=& s_j^z\quad \sigma_i^z(t)\sigma_j^z(0)\sigma_i^z(t)\quad|\begin{tikzpicture}
    % Draw the oval next to the rectangle
    \draw[fill=red!25] (0.1, 0.15) ellipse (0.5 and 0.15);
    % Draw the rectangle
    \draw[fill=blue!25] (0.6, -0.1) rectangle (1.6, 0.4);
    \draw[fill=red!25] (2.1, 0.15) ellipse (0.5 and 0.15);
    \end{tikzpicture}\rangle\nonumber\\
    &=& s_j^z\quad \sigma_i^z(t)\sigma_j^z(0)\quad|\begin{tikzpicture}
    % Draw the oval next to the rectangle
    \draw[fill=red!25] (0.1, 0.15) ellipse (0.5 and 0.15);
    \end{tikzpicture}
    % Draw the rectangle
    \Big(\begin{tikzpicture}
        \draw[fill=blue!25] (0.75, -1) rectangle (1.6, -0.6);
    \end{tikzpicture}_1+
    \begin{tikzpicture}
        \draw[fill=blue!25] (2.5, -1) rectangle (1.6, -0.6);
    \end{tikzpicture}_2+
    \begin{tikzpicture}
        \draw[fill=blue!25] (2.6, -1) rectangle (1.6, -0.6);
    \end{tikzpicture}_3+\ldots
    \Big)
    \begin{tikzpicture}
    \draw[fill=red!25] (5.1, 0.15) ellipse (0.5 and 0.15);
    \end{tikzpicture}\rangle\nonumber\\
    &=& (s_j^z)^2 \quad \sigma_i^z(t)\quad|\begin{tikzpicture}
    % Draw the oval next to the rectangle
    \draw[fill=red!25] (0.1, 0.15) ellipse (0.5 and 0.15);
    \end{tikzpicture}
    % Draw the rectangle
    \Big(\begin{tikzpicture}
        \draw[fill=blue!25] (0.75, -1) rectangle (1.6, -0.6);
    \end{tikzpicture}_1+
    \begin{tikzpicture}
        \draw[fill=blue!25] (2.5, -1) rectangle (1.6, -0.6);
    \end{tikzpicture}_2+
    \begin{tikzpicture}
        \draw[fill=blue!25] (2.6, -1) rectangle (1.6, -0.6);
    \end{tikzpicture}_3+\ldots
    \Big)
    \begin{tikzpicture}
    \draw[fill=red!25] (5.1, 0.15) ellipse (0.5 and 0.15);
    \end{tikzpicture}\rangle\nonumber\\
    &=& (s_j^z)^2 [\sigma_i^z(t)]^2 \quad|\begin{tikzpicture}
    % Draw the oval next to the rectangle
    \draw[fill=red!25] (0.1, 0.15) ellipse (0.5 and 0.15);
    % Draw the rectangle
    \draw[fill=blue!25] (0.6, -0.1) rectangle (1.6, 0.4);
    \draw[fill=red!25] (2.1, 0.15) ellipse (0.5 and 0.15);
    \end{tikzpicture}\rangle = |\psi\rangle,
    \label{eq:otoc_symbol}
\eea
where we have used $\sigma_j^z(0)|\psi\rangle=s_j^z|\psi\rangle$, $(\sigma_i^z(t))^2=\mathds 1$ for all time $t$, and the rectangles with subscripts denote the different configurations to which the initial active block is connected under the action of $\ham_F^{(1)}$. From Eq. (\ref{eq:otoc_symbol}), it can be readily seen that $F(r,t)=1$ when $i$ is in the active block and $j$ is in the inactive block, implying that no local information can propagate to the inactive blocks. 

This same construction hinders `communication' between two active blocks separated by an inactive block, which essentially hints at spatial localization of quantum information. This can be seen by considering a state of the form $|\psi\rangle=|\begin{tikzpicture}
    % Draw the oval next to the rectangle
    \draw[fill=blue!25] (0.1, -0.5) rectangle (1.2, -0.1);
    % Draw the rectangle
    \draw[fill=red!25] (1.7, -0.23) ellipse (0.5 and 0.22);
    \draw[fill=blue!25] (3.2, -0.5) rectangle (2.2, -0.1);
\end{tikzpicture}\rangle$ and consider two sites $i$, lying within the left active block (rectangular region) and $j$, lying within the right active block. The most important thing to notice here is that under the action of $\ham_F^{(1)}$, the evolution of the right block happens independently of that of the left active block, i.e.,
\beq
    \ham_F^{(1)}|\psi\rangle=\Big|\Big(\begin{tikzpicture}
        \draw[fill=blue!25] (0.75, -1) rectangle (1.6, -0.6);
    \end{tikzpicture}_1+
    \begin{tikzpicture}
        \draw[fill=blue!25] (2.5, -1) rectangle (1.6, -0.6);
    \end{tikzpicture}_2+
    \begin{tikzpicture}
        \draw[fill=blue!25] (2.6, -1) rectangle (1.6, -0.6);
    \end{tikzpicture}_3+\ldots
    \Big) 
    \begin{tikzpicture}
    \draw[fill=red!25] (5.1, 0.15) ellipse (0.5 and 0.15);
    \end{tikzpicture}
    \Big(\begin{tikzpicture}
        \draw[fill=blue!25] (0.75, -1) rectangle (1.6, -0.6);
    \end{tikzpicture}_1+
    \begin{tikzpicture}
        \draw[fill=blue!25] (2.5, -1) rectangle (1.6, -0.6);
    \end{tikzpicture}_2+
    \begin{tikzpicture}
        \draw[fill=blue!25] (2.6, -1) rectangle (1.6, -0.6);
    \end{tikzpicture}_3+\ldots
    \Big) \Big\rangle.
    \label{eq:otoc_symbol_2}
\eeq
This is a consequence of the non-melting of the central inactive block. Using this property, it can be immediately seen that in this case, too, 
\beq
    \sigma_i^z(t)\sigma_j^z(0)\sigma_i^z(t)\sigma_j^z(0)|\psi\rangle = \left(s_j^z\right)^2 [\sigma_i^z(t)]^2 |\psi\rangle = |\psi\rangle,
\eeq
which implies that $F(r,t)=1$, indicating that each active block is oblivious to the presence of another active block as far as local information is concerned. These features are evident in Fig. 3(a) of the main text.

\subsection{Mutual Information}
\label{MIsection}
\begin{figure}
    \centering
    \includegraphics[width=1.0\linewidth]{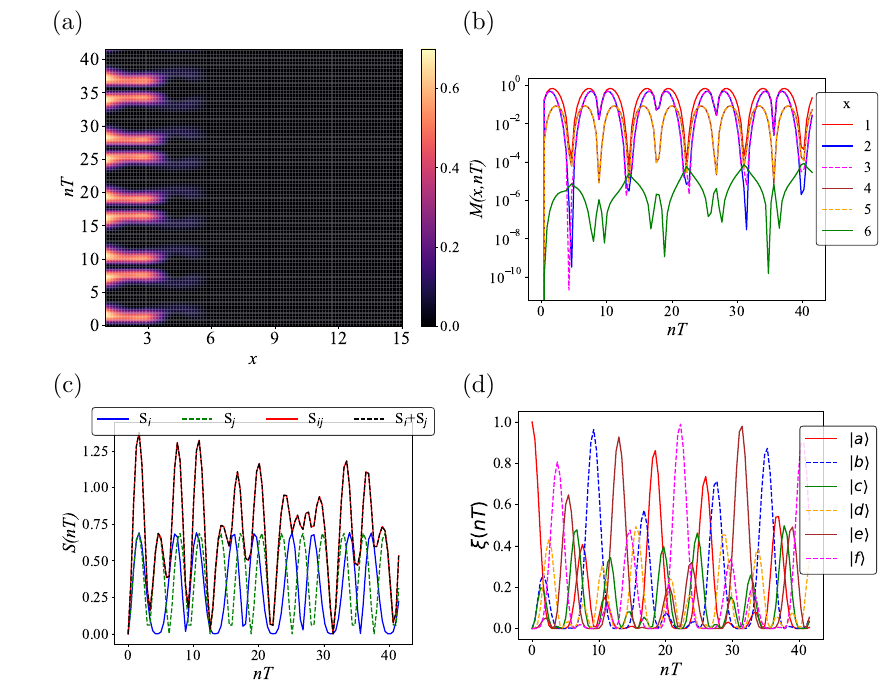}
    \caption{(Color online) (a) Time evolution of the mutual information $M(x,nT)$ with $|a\rangle$ as the initial state (refer text) at $\gamma_0=2\pi$. This state has two active blocks spanning from sites $1-4$ and $11-16$ respectively separated by an inactive block. The spin at site $i=16$ is only entangled with spins at sites $11-15$, which results in $M(x,nT)=0$ beyond $x=5$ (i.e. beyond $j=11$). (b) A 2d log-scale plot of $M(x,nT)$ for $x=1-6$ showing the periodic dips. (c) Illustrative plot for the time evolution of $\mathcal{S}_{16}$ (blue solid), $\mathcal{S}_j$ ($j$ chosen to lie within the second active block, $j=1$) (green dashed), $\mathcal{S}_{16}+\mathcal{S}_1$ (black dashed) and $\mathcal{S}_{1,16}$ (red solid). The black dashed and the red solid curves overlap, confirming that $\mathcal{S}_{ij}=\mathcal{S}_i+\mathcal{S}_j$ for unentangled spins. (d) Plot of the evolution of the overlap $|\langle \phi|\psi(nT)\rangle|^2$ with the six states lying in the same fragment as the initial state. The time evolutions are carried using exact Floquet unitary and $V_0=20$. See text for details.}
    \label{fig:sfig4}
\end{figure}

Another quantity, which is expected to capture the spatial localization of quantum information in this model, is the mutual information between two subsystems, defined as
\beq
    M(x,nT)= \mathcal{S}_i (nT)+ \mathcal{S}_j(nT) - \mathcal{S}_{ij}(nT),
    \label{eq:MI_defn}
\eeq
where $\mathcal{S}_i$ and $\mathcal{S}_j$ denote the entanglement entropies of two subsystems $i$ and $j$, separated by a distance $x=|j-i|$. $\mathcal{S}_{ij}\equiv \mathcal{S}(i \cup j)$ is the entanglement entropy of the combined subsystem, $i\cup j$. By virtue of the triangle inequality of the entanglement entropy ($\mathcal{S}_i+\mathcal{S}_j\geq \mathcal{S}_{ij}$), the quantity $M(x,nT)$ is positive semidefinite. When these two subsystems are not entangled with each other, i.e., the reduced density matrix $\rho_{ij}=\rho_i \bigotimes \rho_j$, then $\mathcal{S}_{ij}=\mathcal{S}_i+\mathcal{S}_j$, thus giving $M(x,nT)=0$.

It can be readily verified from Eq. (\ref{eq:otoc_symbol_2}) that if $i$ and $j$ are chosen to lie on the left and right active blocks, respectively, then $\rho_{ij} (nT)$ decomposes into $\rho_i (nT)\bigotimes \rho_j(nT)$, yielding $M(x,nT)=0$. The same is true if one of $i$ and $j$ is chosen in the active block and the other in the inactive block.

Figure \ref{fig:sfig4}(a) shows the spatiotemporal profile of $M(x,nT)$ at $\gamma_0=2\pi$, when $i$ and $j$ are chosen to be single-site spins, with $i=16$ and $j$ varied along the chain so that $x=16-j$. The chosen initial state is $|\psi(0)\rangle=|a\rangle=|+--+---+++-+-++- \rangle$; the same as the one chosen in the main text, which exhibits spatial localization under $\ham_F^{(1),e}$. From the plot, it is evident that the spin at the site $16$ is not entangled to any spin beyond the site $11$ up to the first $100$ drive cycles (we checked up to $500$ drive cycles). This is consistent with the fact that the state $|a\rangle $ has an active block from sites $11-16$, followed by an inactive block that separates another active block from sites $1-4$. A 2d plot of the same is presented in Fig. \ref{fig:sfig4}(b) for $x=1-6$, which correspond to sites $j=15-10$ respectively. An oscillatory behavior of $M(x,nT)$ is seen, which is explained below. In addition, it is observed that for $x=6$ (i.e., $j=10$), $M(x,nT)$ is very small but not exactly zero. This is seen to be the case for all $x\geq 6$ (seen in Fig. \ref{fig:sfig4}(a)). Ideally, for evolution only with $\ham_F^{(1),e}$, these values would have been exactly zero; however, due to the presence of higher-order terms in the perturbation series, there is some propagation within the inactive block, which is suppressed as the drive frequency is increased. In Fig. \ref{fig:sfig4}(c), we plot the evolutions of the entanglement entropies for $i=16$, a representative site $j=1$ (across the inactive block), the combined subsystem ($1\cup 16$) and the bare sum $\mathcal{S}_1+\mathcal{S}_{16}$. It can be seen that, indeed, for this pair of sites, $\mathcal{S}_i+\mathcal{S}_j=\mathcal{S}_{ij}$, highlighting the product structure of the state as expected. 

To understand the oscillatory behavior of $M(x,nT)$ for $x<6$, we focus on other states belonging to the same fragment as $|a\rangle$ and compute their overlaps with the state $|\psi(nT)\rangle$, $\xi(nT)=|\langle \phi|\psi(nT)\rangle|^2$ in Fig. \ref{fig:sfig4}(d). There are $6$ states in the fragment which hosts $|a\rangle$, namely
\bea
    |\phi\rangle = \{&|a\rangle:|+--+---+++-+-++- \rangle,  &|b\rangle: |- + + - - - - + + + - + - + + -\rangle, \nonumber\\ 
    &|c\rangle : |+ - - + - - - + + + - + + - - +\rangle,  
    &|d\rangle: |- + + - - - - + + + - + + - - +\rangle, \nonumber\\
    &|e\rangle: |+ - - + - - - + + + + - - + - +\rangle, 
    &|f\rangle: |- + + - - - - + + + + - - + - +\rangle \}.
    \label{eq:frag_states}
\eea
We focus at $nT\sim5$ when the first dip in $M(x,nT)$ occurs in Fig \ref{fig:sfig4}(b). The state at this time reads $|\psi(5)\rangle \approx \lambda_1 |d\rangle + \lambda_2|e\rangle + \lambda_3|f\rangle$ with $\lambda_3 > \lambda_1,\lambda_2$. The spin at the last site is seen to be in a direct product state with the rest of the chain so that $S_{16}(5)\sim 0$, thereby giving $M(x,5)\sim 0$ for all $x$. The state near $nT\sim 10$ is $|\psi(10)\rangle \approx \mu_1 |a\rangle + \mu_2 |b\rangle$ with $\mu_2 \gg \mu_1$. Here, too the last spin is almost non-entangled with the rest of the chain resulting in $M(x,10) \sim 0$. Around $nT\sim 20$, the state reads $|\psi(20)\rangle\sim \nu_1 |a\rangle + \nu_2 |b\rangle +\nu_3 |c\rangle$, with $\nu_1> \nu_2,\nu_3$. Writing down the state explicitly shows that the sites $11$ ($x=5$) and $12$ ($x=4$) has a direct product structure with the rest of the chain. Thus $\mathcal{S}_{11}, \mathcal{S}_{12}\sim 0$, giving $M(4,20),M(5,20)\sim 0$. 

The peak near $nT\sim 2$ is reminiscent of a maximally entangled Bell state. The state at this time, $|\psi(2)\rangle\approx \frac{1}{2}(|a\rangle+|b\rangle+|c\rangle+|d\rangle)$. Straightforward calculations give $\mathcal{S}_{16}= \log2,$ $\mathcal{S}_{15}=\log2$, $\mathcal{S}_{14}=\log2$, $\mathcal{S}_{16,15}\sim \log2$ and $S_{16,14}\sim \log2$. This yields $M(1,2), M(2,2) \sim \log2$. We note here that the evolution of the entanglement entropy, in this case, is also very different from what is expected in ergodic systems. In the latter case, the entanglement entropy is expected to saturate to the Page value at long times. However, in this case, because of the presence of a few states in the fragment, the entanglement entropy exhibits an oscillatory behavior as a result of recurrence. 

\subsection{2d plot to examine oscillatory behavior of OTOC in Fig 3(a) of main text}
\begin{figure}
    \centering
    \includegraphics[width=0.5\linewidth]{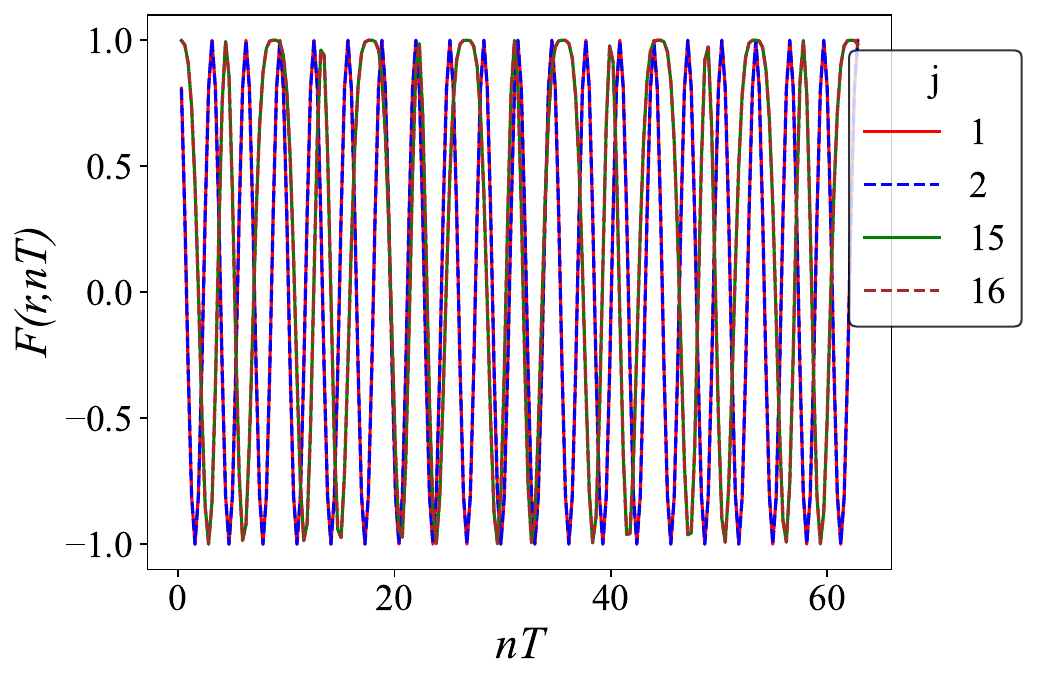}
    \caption{2d plot of $F(r,nT)$ as a function of drive cycle, $n$ for four representative sites $j=1,2,15,16$ when the initial operator is localized at sites $i=2$ and $11$. The oscillations can be clearly seen here. For details, see Fig 3(a) of main text and the discussion in this section.}
    \label{fig:sfig5}
\end{figure}

In this section, we plot (Fig. \ref{fig:sfig5}) a 2d version of the OTOC in Fig. 3(a) of the main text for four representative sites $j=1,2,15,16$. The oscillations seen previously in the density plot are clearer here. For the purpose of recalling, the state chosen was $|a\rangle=|+--+---+++-+-++- \rangle$, which has two active blocks from sites $1-4$ and $11-16$, separated by an inactive region. The initial information is localized at sites $i=2$ and $11$, i.e., $\sigma_i^z(0)\equiv \sigma_2^z\sigma_{11}^z$. This oscillatory behavior can be attributed to the presence of only $6$ states in the fragment of $\ham_F^{(1),e}$ to which $|a\rangle$ belongs. Similar behavior was seen in Ref. \cite{sghosh2}. 

\subsection{A state which exhibits spatial localization under both $\ham_F^{(1),o}$ and $\ham_F^{(1),e}$}
\begin{figure}
    \centering
    \includegraphics[width=1.0\linewidth]{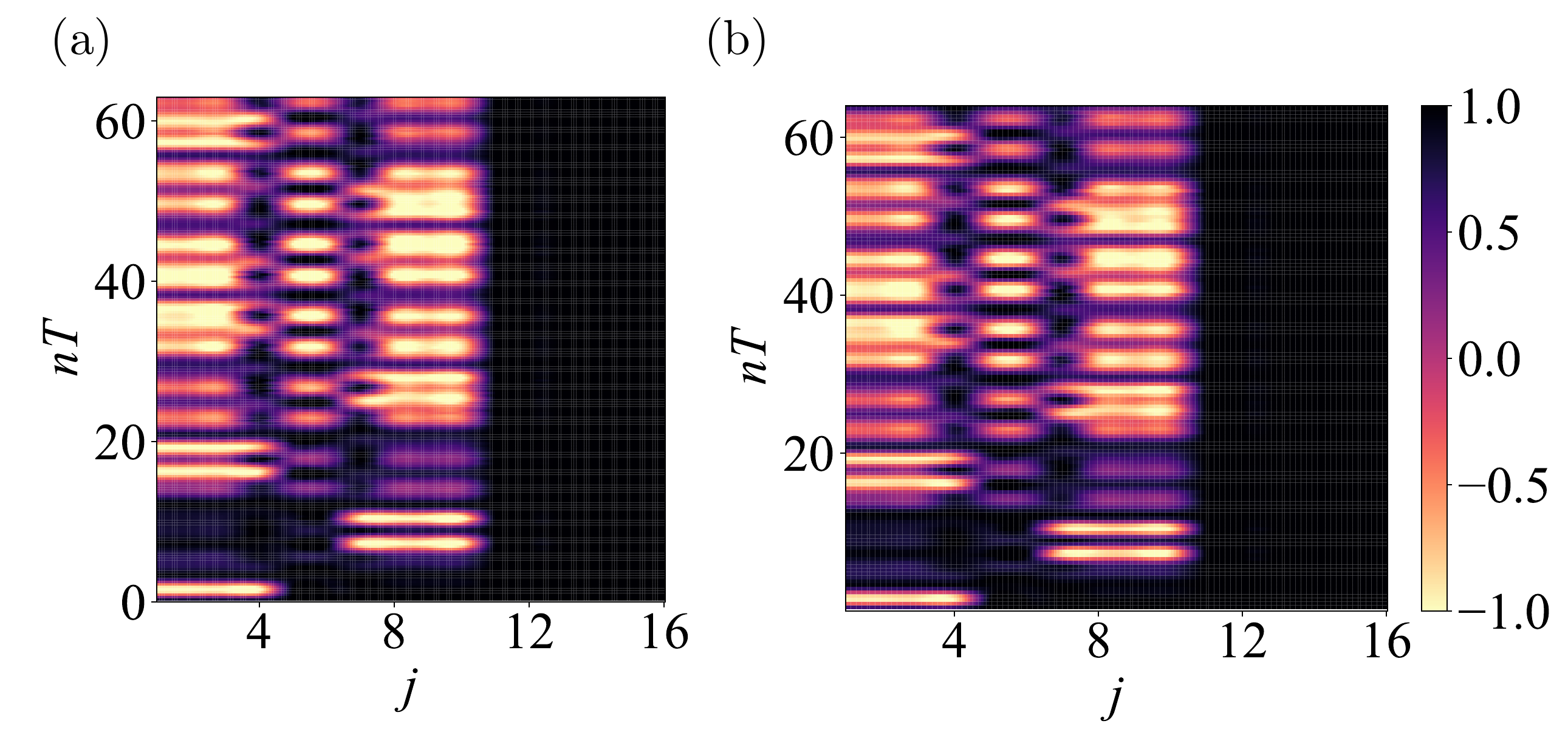}
    \caption{(Color online) Spatio-temporal profile of the OTOC $F(r,nT)$ starting from a state $|\alpha\rangle$ which, with respect to both $\ham_F^{(1),e}$ and $\ham_F^{(1),o}$, has an active block from sites $j=1-10$ and an inactive block from $j=11-16$. The initial information in both the figures is localized at $i=1$, i.e., $\sigma^z_i(0)\equiv \sigma_1^z$. (a) At $\omega^*_e=V_0/\hbar$, the evolution is governed by $\ham_F^{(1),e}$ to leading order and information remains localized within the active region of space. (b) At $\omega_o^*=2V_0/(3\hbar)$ also, when the dynamics is governed by $\ham_F^{(1),o}$, the information remains localized. All time evolutions are carried using exact Floquet operator $\uevol_F(T,0)$ and the drive amplitude $V_0=20$. See text for details.}
    \label{fig:sfig3}
\end{figure}

We study the profile of OTOC $F(r,nT)$ starting from a state $|\alpha\rangle=|- + + - - - - + + - + + + + - - \rangle$, which has an active block from sites $1-10$ and the rest of the chain is an inactive block. Note that in this case, spins in the active region can move only by dipole moment conserving moves; hence the state behaves identically under $\ham_F^{(1),o}$ and $\ham_F^{(1),e}$. In both the cases, information can be localized within the active region. Figures \ref{fig:sfig3}(a) and \ref{fig:sfig3}(b) sketch the profile of $F(r,nT)$ when the system is driven at a frequency $\omega^*_e=V_0/\hbar$ and $\omega^*_o=2V_0/(3\hbar)$, respectively, with drive amplitude $V_0=20$. The initial information in both the cases is localized at the leftmost site $i=1$. The localization of information can be clearly seen from the plots.